\documentclass[letterpaper]{article} %

\usepackage[preprint]{aaai2027}  %
\usepackage[hyphens]{url}  %
\usepackage{graphicx} %
\usepackage{natbib}  %
\usepackage{caption} %
\usepackage{amsmath}
\usepackage{amssymb}
\usepackage{booktabs}
\usepackage{xcolor}
\usepackage{pdflscape}
\usepackage{longtable}
\usepackage{array}
\usepackage{multirow}
\usepackage{siunitx}
\newcommand{\enc}{E}                          %
\newcommand{\dec}{D}                          %
\newcommand{\A}{A}                            %
\newcommand{\Ainv}{A^{-1}}                    %
\newcommand{\quant}{q}                        %
\newcommand{\fills}{\varphi}                  %
\newcommand{\keepproj}{\Pi_{z}}               %
\newcommand{\codes}{z}                        %
\newcommand{\qcodes}{z_q}                     %
\newcommand{\free}{f}                         %

\newcommand{\R}{\mathbb{R}}
\newcommand{\fiber}{\mathcal{F}}              %

\newif\ifmushra
\mushratrue  %

\newif\ifswafifty
\swafiftytrue

\newif\ifcameraready
\camerareadytrue

\newif\ifappendsupplement
\appendsupplementtrue

\title{LILAC: An Idempotent Neural Speech Codec}
\author{June Young Yi\textsuperscript{\rm 1}, Dongwook Lee\textsuperscript{\rm 2}, Jiheum Yeom\textsuperscript{\rm 3}, Sungroh Yoon\textsuperscript{\rm 2,3,4\,$\dagger$}}
\affiliations{\textsuperscript{\rm 1}Department of Computer Science and Engineering, Seoul National University\\
\textsuperscript{\rm 2}Interdisciplinary Program in Artificial Intelligence, Seoul National University\\
\textsuperscript{\rm 3}Department of Electrical and Computer Engineering, Seoul National University\\
\textsuperscript{\rm 4}AIIS, ASRI, INMC, and ISRC, Seoul National University\\
\{julianyi1, dwsmart32, quilava1234, sryoon\}@snu.ac.kr}

\begin{document}

\maketitle
\begingroup
\renewcommand{\thefootnote}{}
\footnotetext{\hspace{-1.6em}$^\dagger$Correspondence to: Sungroh Yoon \textless sryoon@snu.ac.kr\textgreater.}
\endgroup

\begin{abstract}
  Neural Audio Codecs are widely adopted in speech generation and
  editing. However, existing neural audio codecs are not idempotent:
  across the paper's twelve baseline systems, every configuration
  tested rewrites, on average, at least \SI{15}{\%} of its tokens in
  a single decode--re-encode pass. This poses a problem for utilizing
  Neural Audio Codecs as token interfaces in pipelines where
  re-encoding decoded outputs can occur. We present LILAC, a fully
  convolutional \SI{24}{\kilo\hertz} speech codec at \SI{9.375}{\hertz} and
  \SI{0.75}{\kilo\bit\per\second} that is codec idempotent by
  construction; re-encoding the decoded audio of any valid token
  stream returns the identical stream. LILAC achieves idempotency
  while maintaining competitive quality, reaching UTMOS 4.14 and 4.24
  on LibriSpeech and LibriTTS-R test sets, comparable to SOTA
  sub-\SI{1}{\kilo\bit\per\second} Neural Audio Codecs.
\end{abstract}

\section{Introduction}

Neural speech codecs convert audio signals into latent
representations and their corresponding discretized token indices.
Their encoder, quantization bottleneck, and decoder are jointly
trained with a one-pass reconstruction objective
\citep{zeghidour2021soundstream,defossez2023encodec,kumar2023dac}.
However, this
objective does not require a token stream to stay unchanged when
decoded and re-encoded. This omission raises an issue for pipelines
that generate, edit, store, or retransmit these tokens.

\begin{figure*}[t]
  \centering
  \includegraphics[width=0.98\textwidth]{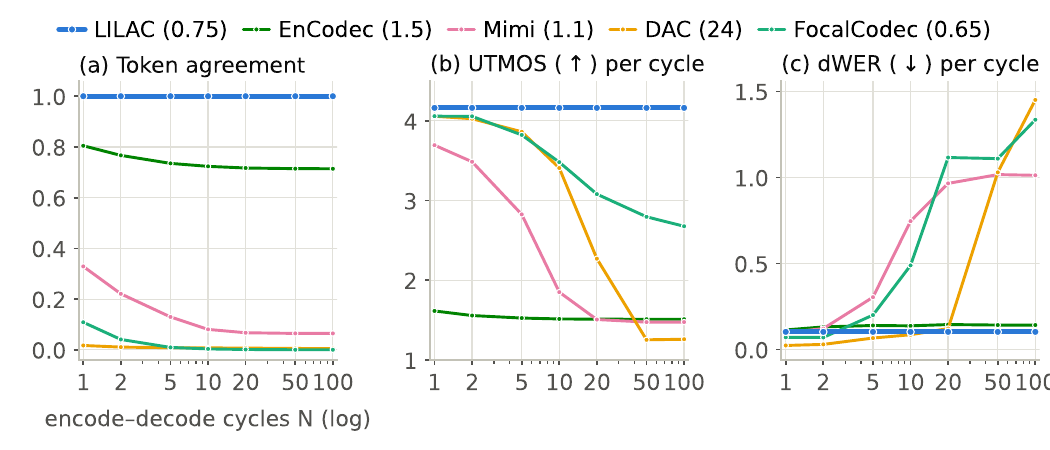}
  \caption{Iterated re-encoding on 100 reader-balanced LibriSpeech
    test-clean clips: token agreement with the first encoding, UTMOS, and
  Whisper dWER over $N=1$--$100$ cycles. Parenthesized rates in kb/s.}
  \label{fig:drift-hook}
\end{figure*}

This decode--re-encode round-trip drift affects almost all existing
codecs. Previous research \citep{oreilly2025codedrift} attempted to
reduce code drift by fine-tuning with specialized losses, but the
auxiliary losses dampened the drift without eliminating it.
Any residual error can still accumulate and cause the token stream to
diverge after repeated cycles. Figure~\ref{fig:drift-hook} demonstrates
various codecs all degrading over 100 decode--re-encode cycles. As
errors accumulate, intelligibility and naturalness both suffer. This
impacts both low-bitrate and high-bitrate speech codecs.

LILAC (Lifting-Inspired Low-rate Audio Codec) is a fully
convolutional, full-band (\SI{24}{\kilo\hertz}),
\SI{0.75}{\kilo\bit\per\second} speech codec which operates at
\SI{9.375}{\hertz} with 80 bits per frame. Its encoder transforms the
original audio signal through a series of invertible transforms and
selectively discards information. We employ Finite Scalar
Quantization (FSQ)~\citep{mentzer2023fsq} to quantize the
coordinates. The decoder then reconstructs the discarded information
from the transmitted quantized coordinates and applies the exact
inverse transform. This guarantees that re-encoding only discards
this reconstructed information, making a decode--re-encode cycle
idempotent. This is independent of model convergence, reconstruction
quality, or any individual weights.

To the best of our knowledge, LILAC is the first neural speech codec
that guarantees codec idempotence by construction. We supply an
algebraic proof of this idempotence guarantee, a conv-only
architecture utilizing finite scalar quantization, and comparisons
with other sub-\SI{1}{\kilo\bit\per\second} codecs. Code, the trained
checkpoint, and an audio demonstration page are publicly
available.\footnote{Code: \url{https://github.com/Rick-McCoy/lilac-codec};
checkpoint: \url{https://huggingface.co/julianyi1/lilac}; audio
demonstrations: \url{https://rick-mccoy.github.io/lilac-demo}.}

\section{Background and Related Work}

\paragraph{Neural audio codecs.}
SoundStream~\citep{zeghidour2021soundstream},
Encodec~\citep{defossez2023encodec}, and DAC~\citep{kumar2023dac}
established the commonly used template
of encoder, discrete bottleneck, and learned decoder. Recently,
low-framerate and low-bitrate codecs have been developed for more
efficient usage in quadratic-compute algorithms such as attention.
These include WavTokenizer~\citep{ji2024wavtokenizer},
BigCodec~\citep{xin2024bigcodec}, SNAC~\citep{siuzdak2024snac},
Mimi~\citep{defossez2024moshi},
SpeechTokenizer~\citep{zhang2023speechtokenizer},
and LFSC~\citep{casanova2024lowframerate}. These systems reach
near-\SI{1}{\kilo\bit\per\second} with
a variety of methods, usually relying on a large decoder to
synthesize discarded details. They optimize for encode--decode
reconstruction without concern for decode--re-encode cycles.

\paragraph{Invertible transforms.}
LILAC employs a series of invertible transforms as its core backbone.
The primary components are additive coupling and learned invertible
$1\times1$ convolutions, standard normalizing-flow
primitives~\citep{papamakarios2021normalizing}. Additive couplings
originate from NICE~\citep{dinh2014nice} and are used in a variety of
other tasks, such as invertible convolutional networks and
transformers~\citep{gomez2017reversible,kitaev2020reformer}. These
also relate to the lifting steps of wavelet
transforms~\citep{sweldens1998lifting}, from which LILAC takes its
name. Invertible $1\times1$ convolutions come from
Glow~\citep{kingma2018glow}, which parametrizes general invertible
kernels. We further constrain the kernel to be orthogonal so that
inversion is equivalent to transposition. Combined with additive
couplings, whose inversion is exactly subtraction, the inverse
transform requires no division. This prevents floating-point
overflows. These operations combine into a volume-preserving transform.

\paragraph{Idempotence.}
Previous research \citep{oreilly2025codedrift} recognized that
repeated decode--re-encode cycles degraded codec reconstructions to
varying degrees and attempted to mitigate it by fine-tuning with
specialized loss functions. \citet{yoshimura2018lossless} constructed
a lossless neural speech codec which is trivially both audio- and
codec-idempotent, but its bitrate is far above what can be utilized
in downstream tasks. Outside the neural speech codec domain, the
Idempotent Generative Network~\citep{shocher2024idempotent} optimizes
for idempotence also with specialized loss functions. LILAC instead
guarantees codec idempotence by construction.

\section{Codec Idempotence}

A neural speech codec consists of an encoder $\enc:X\to C$ and
decoder $\dec:C\to X$. It is \textbf{codec idempotent} when
\[
  \enc(\dec(c)) = c \qquad \text{for every } c \in C.
\]
Note that codec idempotence is separate from \textbf{audio
idempotence}, which is when
\[
  \dec(\enc(x)) = x \qquad \text{for every } x \in X.
\]
While audio idempotence is achieved in previous research such as
\citep{yoshimura2018lossless}, it requires a
prohibitively large codec space in which
audio-to-codec encoding must be injective. Instead, codec idempotence
matches a codec $c$ with its fiber $\fiber_c=\{x\in X:\enc(x)=c\}$.
An idempotent decoder then needs to map $c$ to a waveform in
$\fiber_c$. Training determines the shape of $\fiber_c$ and the
waveform it selects.

\paragraph{Construction.}
Let $\A: \R^{T} \to \R^{T}, \A(x) = (\codes, \free)$ be the
invertible analysis transform, where $\codes \in \R^{d}$ are retained
coordinates and $\free \in \R^{T-d}$ are discarded coordinates. Let
$\quant$ be a finite scalar quantizer. Define
\[
  \enc(x) = \quant \left(\keepproj \A(x) \right), \quad
  \dec(\qcodes) = \Ainv \left(\qcodes, \fills(\qcodes) \right)
\]
where $\keepproj$ selects retained coordinates and the learned fill
$\fills$ predicts the discarded coordinates.

\paragraph{Proposition (idempotence).}
For every $\qcodes$ in the image of $\quant$ and every fill $\fills$,
$\enc(\dec(\qcodes)) = \qcodes$ holds regardless of weights.

\emph{Proof.} Invertibility gives $\A(\dec(\qcodes)) = (\qcodes,
\fills(\qcodes))$, hence $\keepproj \A(\dec(\qcodes)) = \qcodes$.
Finite scalar quantization consists of clamping and rounding each
coordinate. Since clamping and rounding are both idempotent, $\quant$
is also idempotent. Thus,
\[
  \enc(\dec(\qcodes))
  = \quant \left(\keepproj \A(\dec(\qcodes)) \right)
  = \quant (\qcodes) = \qcodes. \quad \blacksquare
\]

The proof requires no constraints on $\fills$. This means that we can
construct a stochastic decoder $\fills(\qcodes, \omega)$ with an
auxiliary input $\omega$ which also shares the same idempotence guarantee.

\section{Method}

LILAC consists of two components: the invertible analysis transform
$\A$ and the fill $\fills$ which predicts discarded coordinates. Both
modules are fully convolutional.

\begin{figure*}[t]
  \centering
  \includegraphics[width=\textwidth]{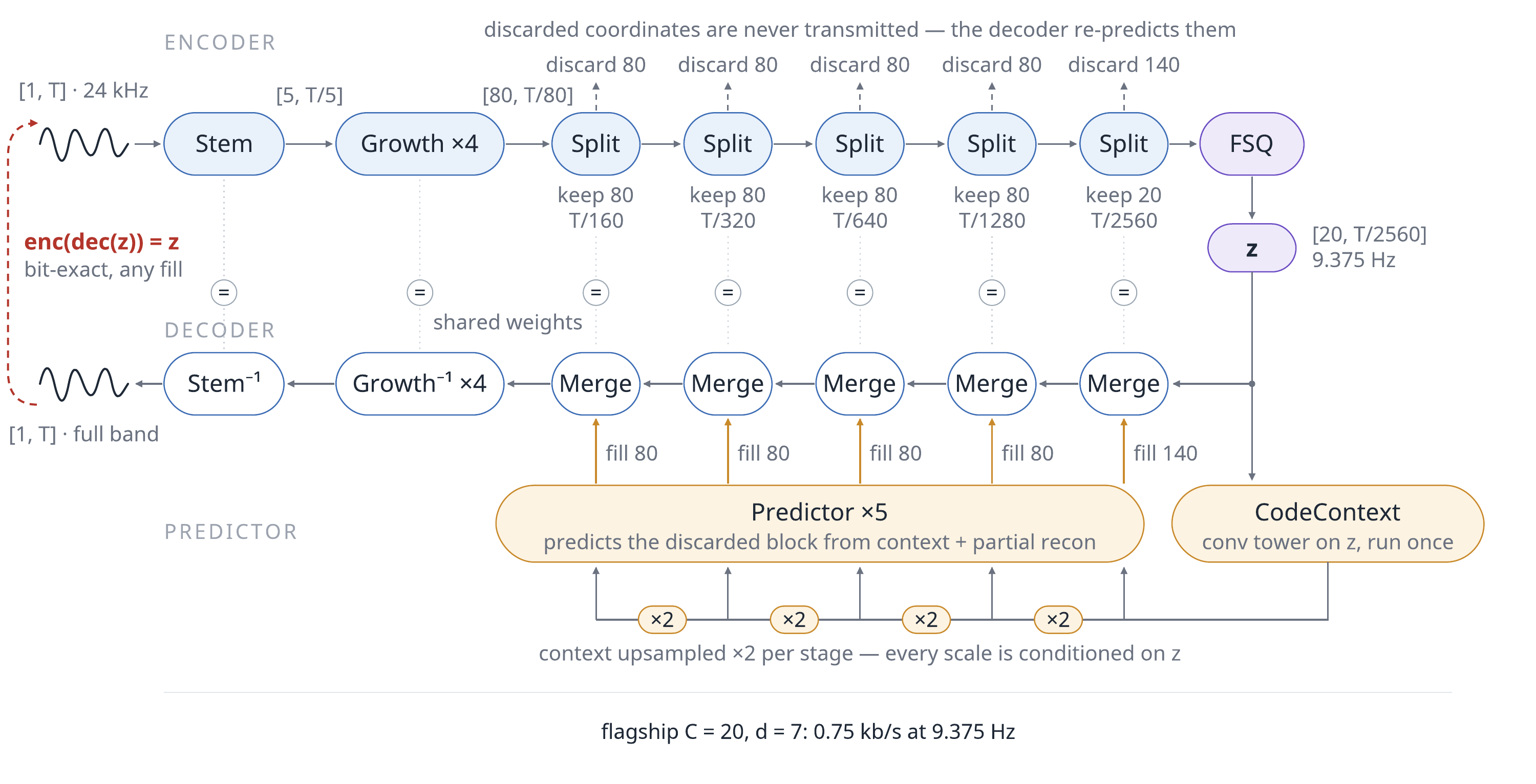}
  \caption{LILAC's encoder (top), shared-weight inverse (middle), and
    decoder-only fill networks (bottom). Projection stages transmit the
    retained coordinates and discard the rest; the decoder predicts the
    discarded blocks before inversion. Re-encoding recovers $z_q$ and drops
    those predictions, closing the exact red loop for any fill. The
  flagship uses $C=20,d=7$.}
  \label{fig:architecture}
\end{figure*}

\subsection{Analysis Transform}

There are two main blocks that make up the Analysis Transform -- the
\textbf{Invertible $1 \times 1$ Convolution} and the \textbf{Additive
Coupling Block}.

The \textbf{Invertible $1 \times 1$ Convolution} is a convolution
with an orthogonal weight kernel and no bias. The input and output
channel counts are the same, and we apply an orthogonal
parametrization to the weight kernel. Inverting the block simply
requires performing a convolution with the weight kernel transposed.
The name $1 \times 1$ comes from the original
literature~\citep{kingma2018glow} where it deals with 2-dimensional
images and thus uses 2-dimensional kernels. Although the audio data
type is 1-dimensional and thus we use 1-dimensional kernels, we
retain the name $1 \times 1$ for familiarity.

The \textbf{Additive Coupling Block} is comprised of three components
-- an input $1 \times 1$ convolution, the additive coupling operation,
and an output $1 \times 1$ convolution. The additive coupling
operation operates on a simple principle: given an input $x$ and two
functions $f,g$, we can split $x$ into $x_{1},x_{2}$ and do the following:
\[
  y_{1} = x_{2} + f(x_{1}), \quad
  y_{2} = x_{1} + g(y_{1}), \quad
  y = [y_{1}, y_{2}]
\]

One can see that since both $x_{1} = y_{2} - g(y_{1})$ and $x_{2} =
y_{1} - f(x_{1})$ are recoverable from $y$, the additive coupling
operation is invertible.

We use stacked ConvNext1D~\citep{liu2022convnext} blocks for $f$ and
$g$, while splitting $x$ into $x_{1}$ and $x_{2}$ by selecting even
and odd timewise indices. Note that we do not need to recreate the
timewise selection for combining $[y_{1}, y_{2}]$ -- indeed, by
concatenating them channelwise, we manage to halve time and double
channel count. This performs as a squeeze-and-mix stage.

Before these squeeze-and-mix stages, we reshape the $[1,T]$ audio
signal into a $[5,T/5]$ tensor and insert a \textbf{Stem} stage. This
is to ameliorate any polyphase artifacts from the reshape. The Stem
stage consists of an invertible $1 \times 1$ convolution and a
special affine version of the additive coupling block. Here, the
5-channel tensor is split into 2-channel and 3-channel tensors. A
small convolutional network produces a log-scale and bias for the
3-channel tensor from the 2-channel tensor:
\[
  [x_{:2},x_{2:}] = x, [\log s, b] = f(x_{:2}), y_{1} = x_{:2}, y_{2}
  = s \cdot x_{2:} + b
\]

We employ 4 squeeze-and-mix stages to grow 5 channels to 80. We then
continue to apply the same block to halve the time, but this time we
do not transmit the full doubled channel count -- instead, we slice
off a portion of the channels to discard the coordinates. We discard
$80,80,80,80,140$ channels while halving the time each stage,
resulting in a $[20,T/2560]$ tensor.

\subsection{Finite Scalar Quantization}

We allocate 4 bits per channel in the range of $[-1, 1]$, resulting
in a grid of $[-1, -13/15, ..., 13/15, 1]$. This results in 80 bits
per frame. Since the original audio stream is in \SI{24}{\kilo\hertz}
and our hop length is 2560, our frame rate is \SI{9.375}{\hertz} and
our bitrate is \SI{0.75}{\kilo\bit\per\second}.

\subsection{Discarded-Coordinate Fill}

In order to decode this quantized token stream, we run the analysis
transform in reverse while recreating the discarded coordinates. A
convolutional \textbf{code-context network} processes the transmitted
code to produce a context tensor, and a \textbf{context ladder}
upsamples said representation for every projection stage. A
stage-specific residual predictor combines the partial reconstruction
and generated context for inversion.

\subsection{Loss functions}

We follow standard neural-vocoder practice~\citep{kong2020hifigan,defossez2023encodec}
utilized in many codec training recipes: a multi-resolution mel loss,
a multi-resolution STFT loss, an adversarial hinge loss against a
multi-resolution STFT \& multi-period discriminator, and a feature
matching loss, with weights 15, 1, 1, and 2 respectively. Adversarial
and feature-matching terms activate after a \SI{5000}{}-step warm-up.
STFT resolutions, periods, and other discriminator configurations are
available in the supplement.

\section{Experiments}

\subsection{Datasets}

\paragraph{Training.}
We train LILAC on HiFiTTS-2~\citep{langman2025hifitts2}, a
\SI{31700}{\hour} LibriVox-derived dataset with \SI{4629}{} speakers,
resampled from \SI{44.1}{\kilo\hertz} to \SI{24}{\kilo\hertz}. We exclude all 146 LibriSpeech/LibriTTS(-R)
readers with a reader-ID filter. We also separate 40 randomly
selected readers to act as a test set.

\paragraph{Evaluation.}
We evaluate against LibriSpeech test-clean
\citep{panayotov2015librispeech} ($n=\SI{2620}{}$,
\SI{16}{\kilo\hertz}), LibriTTS-R test \citep{koizumi2023librittsr}
($n=\SI{4837}{}$, \SI{24}{\kilo\hertz}), VCTK
\citep{yamagishi2019vctk} ($n=\SI{3094}{}$, \SI{24}{\kilo\hertz}),
and the above-mentioned HiFiTTS-2 test set ($n=585$). LibriSpeech,
LibriTTS-R, and HiFiTTS-2 are all LibriVox-derived, while VCTK is
entirely disjoint.

\subsection{Baselines}

We compare our results against 4 neural speech codecs:
WavTokenizer~\citep{ji2024wavtokenizer},
SNAC~\citep{siuzdak2024snac}, Mimi~\citep{defossez2024moshi}, and
FocalCodec~\citep{dellalibera2025focalcodec}. These have been
selected because they operate under the \SI{1}{\kilo\bit\per\second}
limit we target. We evaluate all rate/quantizer count configurations
they support as long as they are under \SI{1}{\kilo\bit\per\second}.
Evaluation results for more codec families can be found in the
supplement. Note that, while LILAC does not train on the evaluation
datasets, test set or otherwise, this is not the case for the comparison codecs.

\subsection{Metrics}

We evaluate LILAC and the comparison codecs on a variety of metrics,
targeting naturalness, intelligibility, fidelity, and waveform
coherence. UTMOS~\citep{saeki2022utmos} measures naturalness.
SCOREQ~\citep{ragano2024scoreq} measures the perceptual quality
degradation of the reconstructed audio. PESQ~\citep{rix2001pesq}
measures perceived audio quality. STOI~\citep{taal2010stoi} measures
intelligibility. dWER uses
Whisper-large-v3~\citep{radford2022whisper} to measure the difference
in word error rate. SI-SNR~\citep{leroux2019sdr} measures waveform
coherence. We also report Mel-Cepstral Distortion, mel-spectrogram
$L_{1}$ distance, and WavLM~\citep{chen2022wavlm} speaker-embedding
cosine similarity.

Idempotence is measured as the token index agreement after a
decode--re-encode cycle. Figure~\ref{fig:drift-hook} applies the
cycle up to 100 times to 100 randomly selected reader-balanced
LibriSpeech clips, comparing each evaluation point with the first
encoding and scoring the reconstructed audio. Evaluation results for
more codecs are available in the supplement.

\subsection{Training Details}

We set the hidden dimension of ConvNext1D layers to 256 and depth 4.
This results in 58.5M parameters overall, with 43.1M in the shared
analysis transform and 15.4M in the fill networks. Note that due to
the encoder and decoder sharing the analysis transform weights, there
are 43.1M parameters in the encoder and 58.5M parameters in the
decoder. We use a TPU v6e-8 for our training hardware. Training runs
in fp32 with the TPU's default bf16 matmul precision, with one
exception: the invertible $1 \times 1$ convolution layers are
executed in the full-precision setting to preserve round-trip
exactness. We use batch size 256 and \SI{25600}{} sample segments.
For the optimizer, we use AdamW~\citep{loshchilov2019adamw} at $2e-4$
learning rate for the generator, and $1e-4$ learning rate for the
discriminator, with exponential decay 0.999996 per step and gradient
clipping at 0.3. We switch off the decay schedule at step 848k. The
final 50k steps train with a constant learning rate of
$5\times10^{-5}$/$2.5\times10^{-5}$. We average the 10 checkpoints
from 889k--898k steps with Stochastic Weight Averaging~\citep{izmailov2018swa}.

\begin{table}[t]
\centering
\small
\setlength{\tabcolsep}{4pt}
\begin{tabular}{@{}lccc@{}}
\toprule
 & UTMOS$\uparrow$ & dWER$\downarrow$ & SCOREQ$\downarrow$ \\
\midrule
LibriSpeech & $4.019 \pm 0.021$ & $0.123 \pm 0.002$ & $0.329 \pm 0.011$ \\
LibriTTS-R  & $4.154 \pm 0.022$ & $0.102 \pm 0.010$ & $0.279 \pm 0.010$ \\
\bottomrule
\end{tabular}
\caption{Seed variance: mean $\pm$ standard deviation over five seeds of
the full recipe, each trained to 600k steps and averaged over its final
ten checkpoints.}
\label{tab:seed-variance}
\end{table}

\subsection{Seed Variance}

We validate seed variance across 5 seeds of the recipe trained to
600k steps. We average the last 10 checkpoints with 1k step spacing
and evaluate on both LibriSpeech and LibriTTS-R. Results are
available in Table~\ref{tab:seed-variance}; per-seed values are
available in the supplement. As the observed spread is smaller than
the inter-system differences we draw conclusions from, it allays any
concerns about statistical significance.

\section{Results}

\begin{figure*}[t]
  \centering
  \includegraphics[width=0.94\textwidth]{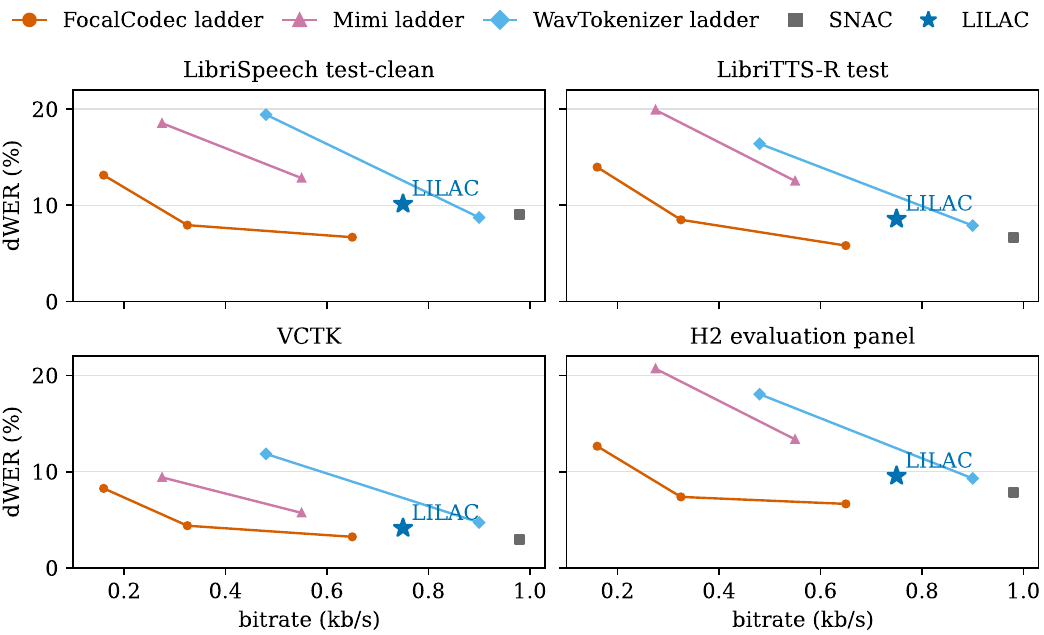}
  \caption{dWER versus released bitrate for the sub-1~kb/s panel; lines
  connect operating points from the same checkpoint family.}
  \label{fig:rate-intelligibility}
\end{figure*}

\subsection{Idempotence Verification}

We empirically verify idempotence holds by decoding and re-encoding
all \SI{7457}{} LibriSpeech and LibriTTS-R test set samples. Every
sample re-encodes to the exact token stream. By selecting 100 random
clips and running 100 cycles, we can compare this with other codecs
as shown in Figure~\ref{fig:drift-hook}. We see that first-pass
agreement is already only 0.805 for EnCodec, 0.329 for Mimi, 0.109
for FocalCodec, and 0.017 for DAC. By cycle 100, FocalCodec and Mimi
have drifted to almost 0 agreement. The same story repeats for
naturalness and intelligibility -- UTMOS and dWER both degrade over
repeated decode--re-encode cycles for every tested codec. A broader
test over 12 codecs confirms that every codec rewrites at least
\SI{15}{\percent} of its tokens on the first pass.

\subsection{Naturalness and Intelligibility at Matched Rate}

LILAC is competitive in naturalness, perceived audio quality,
intelligibility, and waveform coherence. It universally reports the
best score among all comparison codecs in STOI and SI-SNR across all
evaluation datasets. While dWER and SCOREQ are middle-of-the-pack,
UTMOS, Mel-distance, spk-sim, and MCD are all among the top.

Focusing on SI-SNR, we can see that the codecs bifurcate into two
groups: the waveform-coherent and the non-waveform-coherent. This is
because extremely low-rate codecs such as FocalCodec sacrifice
adherence to the original waveform in favor of sounding naturalistic.
This also affects word error rate and intelligibility, as seen in the
high dWER and low PESQ scores of the \SI{12.5}{\hertz} variant of FocalCodec.

\begin{figure*}[t]
  \centering
  \includegraphics[width=0.90\textwidth]{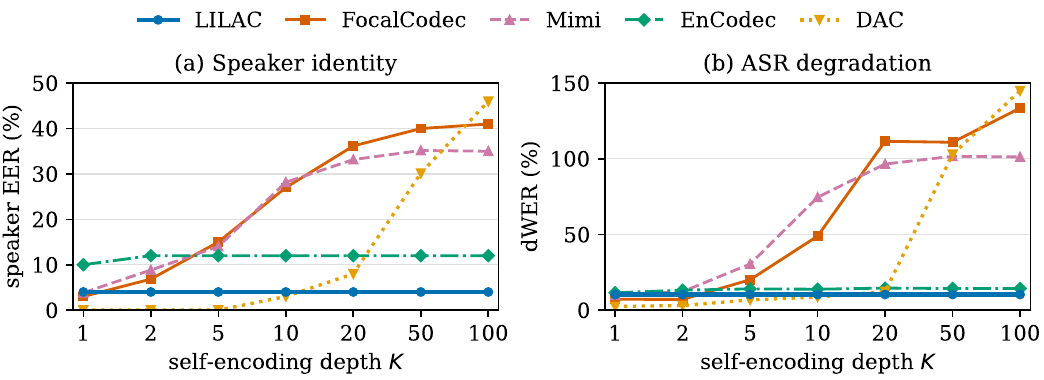}
  \caption{Speaker EER and dWER after each codec self-encodes its own
  output.}
  \label{fig:downstream-k}
\end{figure*}

\subsection{Out-of-Domain Generalization}

Among the three out-of-domain evaluation sets, LILAC scores the best
UTMOS on LibriTTS-R and second-best on VCTK; on LibriSpeech it trails
second place by 0.004, well within seed variance
(Table~\ref{tab:seed-variance}), demonstrating high naturalness even
for out-of-domain datasets. Previous analysis regarding
intelligibility, waveform coherence, and perceived audio quality also
holds against these three datasets.

Note that out of every codec compared, LILAC alone does not contain
these evaluation sets in its training datasets. LILAC is able to
achieve comparable performance despite this handicap. Furthermore,
the degree of domain shift also varies among these evaluation sets.
LibriSpeech and LibriTTS-R are, while speaker-separate from
HiFiTTS-2, still derived from the same LibriVox recording domain.
VCTK is entirely disjoint at the corpus level, making it truly
out-of-domain. LILAC still leads the group in PESQ, STOI, SI-SNR and
is second-best in UTMOS.

\subsection{Downstream Task Stability}

\begin{table*}[t]
  \centering
  \small
  \setlength{\tabcolsep}{2.35pt}
  \begin{tabular}{@{}lrrrrrrrcrrrrrr@{}}
    \toprule
    & & \multicolumn{6}{c}{LibriSpeech test-clean ($n{=}2{,}620$)} & & \multicolumn{6}{c}{LibriTTS-R test ($n{=}4{,}837$)} \\
    \cmidrule(lr){3-8}\cmidrule(lr){10-15}
    Codec & kb/s & UTMOS$\uparrow$ & dWER$\downarrow$ & PESQ$\uparrow$ & SCQ$\downarrow$ & STOI$\uparrow$ & SI-SNR & & UTMOS$\uparrow$ & dWER$\downarrow$ & PESQ$\uparrow$ & SCQ$\downarrow$ & STOI$\uparrow$ & SI-SNR \\
    \midrule
    \emph{Ground truth} & --- & \emph{4.072} & --- & --- & --- & --- & --- & & \emph{4.194} & --- & --- & --- & --- & --- \\
    FocalCodec (12.5~Hz) & 0.16 & \textbf{4.226} & 0.131 & 1.19 & 0.317 & 0.780 & $-37.4$ & & 4.217 & 0.140 & 1.14 & 0.293 & 0.781 & $-35.6$ \\
    Mimi ($n_q{=}2$) & 0.275 & 2.532 & 0.185 & 1.26 & 0.705 & 0.769 & $-24.2$ & & 2.537 & 0.199 & 1.22 & 0.734 & 0.780 & $-16.9$ \\
    FocalCodec (25~Hz) & 0.325 & 4.145 & 0.079 & 1.29 & 0.275 & 0.830 & $-36.1$ & & 4.124 & 0.085 & 1.22 & 0.247 & 0.836 & $-34.4$ \\
    WavTokenizer (40~Hz) & 0.48 & 3.577 & 0.194 & 1.61 & 0.351 & 0.849 & $-30.4$ & & 3.738 & 0.164 & 1.56 & 0.299 & 0.866 & $-27.4$ \\
    Mimi ($n_q{=}4$) & 0.55 & 3.133 & 0.128 & 1.66 & 0.514 & 0.853 & $-5.4$ & & 3.294 & 0.125 & 1.66 & 0.498 & 0.869 & $-1.1$ \\
    FocalCodec (50~Hz) & 0.65 & 4.059 & \textbf{0.067} & 1.39 & \textbf{0.242} & 0.859 & $-34.5$ & & 4.076 & \textbf{0.058} & 1.33 & 0.216 & 0.868 & $-32.9$ \\
    \textbf{LILAC (ours)} & 0.75 & 4.141 & 0.101 & \textbf{2.60} & 0.299 & \textbf{0.935} & $\mathbf{+2.2}$ & & \textbf{4.238} & 0.086 & \textbf{2.60} & 0.252 & \textbf{0.944} & $\mathbf{+6.0}$ \\
    WavTokenizer (75~Hz) & 0.90 & 3.783 & 0.087 & 2.11 & 0.252 & 0.896 & $-1.0$ & & 4.182 & 0.079 & 2.41 & \textbf{0.164} & 0.923 & $+4.6$ \\
    SNAC & 0.98 & 3.276 & 0.090 & 1.97 & 0.453 & 0.889 & $-3.1$ & & 3.951 & 0.067 & 2.24 & 0.192 & 0.918 & $+2.3$ \\
    \midrule
    & & \multicolumn{6}{c}{VCTK ($n{=}3{,}094$)} & & \multicolumn{6}{c}{H2 evaluation panel ($n{=}585$)} \\
    \cmidrule(lr){3-8}\cmidrule(lr){10-15}
    Codec & kb/s & UTMOS$\uparrow$ & dWER$\downarrow$ & PESQ$\uparrow$ & SCQ$\downarrow$ & STOI$\uparrow$ & SI-SNR & & UTMOS$\uparrow$ & dWER$\downarrow$ & PESQ$\uparrow$ & SCQ$\downarrow$ & STOI$\uparrow$ & SI-SNR \\
    \midrule
    \emph{Ground truth} & --- & \emph{4.060} & --- & --- & --- & --- & --- & & --- & --- & --- & --- & --- & --- \\
    FocalCodec (12.5~Hz) & 0.16 & \textbf{4.143} & 0.083 & 1.27 & 0.350 & 0.746 & $-33.8$ & & \textbf{4.162} & 0.126 & 1.20 & 0.290 & 0.781 & $-36.2$ \\
    Mimi ($n_q{=}2$) & 0.275 & 2.858 & 0.094 & 1.40 & 0.694 & 0.722 & $-25.0$ & & 2.423 & 0.207 & 1.24 & 0.684 & 0.766 & $-20.9$ \\
    FocalCodec (25~Hz) & 0.325 & 4.083 & 0.044 & 1.42 & 0.289 & 0.793 & $-32.2$ & & 4.054 & 0.074 & 1.31 & 0.242 & 0.829 & $-35.4$ \\
    WavTokenizer (40~Hz) & 0.48 & 3.769 & 0.118 & 1.63 & 0.337 & 0.806 & $-27.2$ & & 3.557 & 0.181 & 1.61 & 0.274 & 0.843 & $-30.6$ \\
    Mimi ($n_q{=}4$) & 0.55 & 3.345 & 0.057 & 1.73 & 0.547 & 0.787 & $-3.8$ & & 3.024 & 0.134 & 1.64 & 0.469 & 0.850 & $-2.9$ \\
    FocalCodec (50~Hz) & 0.65 & 4.024 & 0.032 & 1.56 & 0.262 & 0.818 & $-31.4$ & & 3.978 & \textbf{0.067} & 1.42 & 0.211 & 0.856 & $-34.0$ \\
    \textbf{LILAC (ours)} & 0.75 & 4.104 & 0.041 & \textbf{2.59} & 0.312 & \textbf{0.867} & $\mathbf{+5.3}$ & & 4.087 & 0.096 & \textbf{2.48} & 0.267 & \textbf{0.929} & $\mathbf{+2.5}$ \\
    WavTokenizer (75~Hz) & 0.90 & 3.587 & 0.047 & 1.82 & 0.270 & 0.835 & $-5.0$ & & 3.642 & 0.093 & 2.01 & \textbf{0.190} & 0.889 & $-0.6$ \\
    SNAC & 0.98 & 3.810 & \textbf{0.029} & 2.29 & \textbf{0.242} & 0.849 & $-1.3$ & & 3.429 & 0.079 & 2.01 & 0.259 & 0.889 & $-1.9$ \\
    \bottomrule
  \end{tabular}
  \caption{Full-split comparison at matched rate ($<$1~kb/s). Bold marks
  the best non-saturated value per column; fidelity panels (MCD,
  mel-$L_1$, speaker similarity) appear in the supplement.}
  \label{tab:main}
\end{table*}

For neural speech codecs, the most common downstream task is for
usage in Text-To-Speech (TTS) models. For TTS models, speaker
identity preservation and linguistic degradation are the two most
important factors. We evaluate Speaker Equal Error Rate (EER) and
dWER on repeated re-encoded samples. Figure~\ref{fig:downstream-k}
and Table~\ref{tab:under-k} show the results: while speaker EER and
dWER start out as the fourth-best among the compared codecs, after
repeated re-encodings, the unchanging LILAC beats all other codecs by
a wide margin.

\begin{table}[t]
\centering
\small
\setlength{\tabcolsep}{3.2pt}
\begin{tabular}{@{}lcccccc@{}}
\toprule
 & \multicolumn{3}{c}{EER (\%)$\downarrow$} & \multicolumn{3}{c}{dWER$\downarrow$} \\
\cmidrule(lr){2-4}\cmidrule(lr){5-7}
Codec & $K{=}1$ & $K{=}10$ & $K{=}100$ & $K{=}1$ & $K{=}10$ & $K{=}100$ \\
\midrule
LILAC & 4.00 & 4.00 & 4.00 & 0.103 & 0.103 & 0.103 \\
FocalCodec & 3.00 & 27.00 & 41.00 & 0.072 & 0.489 & 1.335 \\
Mimi & 3.83 & 28.17 & 35.00 & 0.099 & 0.746 & 1.012 \\
EnCodec & 10.00 & 12.00 & 12.00 & 0.115 & 0.138 & 0.143 \\
DAC & 0.00 & 3.00 & 46.00 & 0.024 & 0.087 & 1.450 \\
\bottomrule
\end{tabular}
\caption{Speaker EER and dWER after $K$ self-encoding cycles
(Figure~\ref{fig:downstream-k}, selected depths). LILAC's values are
identical at every $K$.}
\label{tab:under-k}
\end{table}

\subsection{Human Evaluation}
\label{sec:mushra}

We ran a crowdsourced MUSHRA-style listening test on a paid platform
with a non-expert pool. It is not strictly BS.1534-conformant
(playback conditions are self-selected and unverified), matching
common practice for neural-codec evaluation. Each listener rated all
nine held-out experimental items (three each from LibriSpeech
test-clean, LibriTTS-R test, and VCTK) across eight conditions per
trial: a hidden reference, 3.5 and \SI{7}{\kilo\hertz} low-pass
anchors, LILAC, and four rate-bracketing peers from
Table~\ref{tab:main} (FocalCodec at 0.65, WavTokenizer at 0.90, SNAC
at 0.98, and DualCodec at \SI{1.23}{\kilo\bit\per\second}). All 40
listeners were screened with session completeness, golden
catch-trials, and flatline rejection, and retained. The hidden
reference averaged 90.6, and inter-rater reliability was ICC(2,1) $=0.66$.

While LILAC outperforms its rate-nearest comparison FocalCodec, the
contrast is not statistically significant ($+2.9$ points where the
  \SI{95}{\percent} CI is $[-3.3, +9.0]$, Holm $p=0.077$, sample means
51.4 versus 48.5). The three systems with $1.2$ - $1.6\times$ LILAC's
bitrate all scored higher. In fact, performance largely correlated
with bitrate. LILAC therefore performs in line with its peers while
achieving codec idempotence. A full analysis of results combined with
full protocol disclosures and screening rules is available in the supplement.

\subsection{Ablations}

\paragraph{Dataset and augmentation.}
Before using HiFiTTS-2, we used the training subsets of both
LibriSpeech and LibriTTS-R to train our model. While this resulted in
lower UTMOS for LibriTTS-R, LibriSpeech UTMOS was lower by $\sim$0.3.
dWER, STOI, and other metrics similarly suffered. Furthermore,
out-of-domain evaluations on VCTK were not comparable to the other
codecs. Switching to HiFiTTS-2 helped with OOD generalization. We did
not include both LibriSpeech and LibriTTS-R train sets in our
training dataset so that both remain OOD when evaluated.

In order to improve LibriSpeech performance, we focused on the fact
that LibriSpeech is \SI{16}{\kilo\hertz} while both LibriTTS-R and
HiFiTTS-2 are \SI{24}{\kilo\hertz}. Of note is that LibriTTS-R has
its high-band component synthesized via
Miipher~\citep{koizumi2023miipher,koizumi2023librittsr}. We therefore augmented the
training dataset by either randomly resampling into lower sampling
rates, applying an 8th-order lowpass filter, or attenuating the
high-band by a random amount. We also randomly apply a loudness
scale. Refer to our training dataset implementation for exact
probabilities and values. A comparison between a non-augmented
training run and an augmented training run at 300k steps is available
in the supplement.

\paragraph{Structural limitations and capacity.}
By freezing the final checkpoint and training a separate
discarded-coordinate fill for 200 LibriTTS-R clips, we can find the
optimal reconstructed coordinates to reconstruct the original
waveforms and determine the headroom of the existing fill network. We
were able to lower the median SCOREQ from 0.256 to 0.205, indicating
a better fill network could have improved performance substantially.
However, simply scaling the hidden dimension size or convolutional
stack depth did not yield any meaningful improvements.

Interestingly, the analysis transform's shared weight constraint did
not yield much room for improvement. After untying the shared weights
and training the encoder and decoder weights separately, we did not
observe any meaningful improvement across all metrics and datasets.
This indicates that in order to improve LILAC's performance, we must
tackle the architecture of either the analysis transform or the fill network.

\paragraph{Bit allocation.}
Even within the \SI{1}{\kilo\bit\per\second} constraint and holding
the \SI{0.75}{\kilo\bit\per\second} bitrate fixed, the frame rate and
FSQ bit allocation can be changed. For example, we could train a
\SI{18.75}{\hertz} variant with 2 bits per coordinate, or a
\SI{37.5}{\hertz} variant with Binary Spherical
Quantization~\citep{zhao2024bsq}, allocating just 1 bit per
coordinate. These variants, while slightly improving UTMOS, decimated
dWER. Eliminating quantization and using continuous latents worsened
every metric. Only by increasing the bitrate by increasing channel
count (decreasing the number of discarded coordinates), frame rate,
or bit depth per coordinate did performance improve without
downsides. This suggests our current configuration is optimal within
the constraints.

\section{Limitations}

LILAC's main contribution is its structural guarantee of codec
idempotence. It does not push the pareto frontier on low-frame-rate,
low-bitrate codec performance. On single-pass metrics, it is merely
competitive among its peers, not SOTA, and it lags behind in certain
areas such as dWER. LILAC scores better than its rate-nearest peer
and trails higher-rate systems in human evaluation.  While this model
is conv-only, it does not perform meaningfully faster than other
codecs. On the contrary, due to the abundance of small-channel
convolutions, LILAC's RTF on GPU accelerators lags behind other
codecs in our current implementation. Finally, the anticipated
benefits of codec idempotence, such as the stability of downstream
systems trained on its tokens, are only theoretical; we have yet to
demonstrate this empirically with a trained downstream model.

\section{Conclusion}

To the best of our knowledge, LILAC is the first neural speech codec
that is codec idempotent by construction. This is achieved by
utilizing an invertible analysis transform as the backbone,
discarding pre-selected coordinates, and only synthesizing those
discarded coordinates. The result is a \SI{24}{\kilo\hertz},
\SI{0.75}{\kilo\bit\per\second}, \SI{9.375}{\hertz} codec that
achieves competitive out-of-domain quality compared to other codecs
in the same bitrate region. LILAC is able to preserve naturalness,
waveform coherence, intelligibility, and audio quality through an
arbitrary number of re-encoding cycles. Future work will focus on
improving the SCOREQ/dWER performance and increasing human
preference. As shown in the proof section, a stochastic variant would
also be able to achieve diverse outputs while still maintaining codec
idempotence.

\section{AI Assistance Disclosure}

AI coding agents were used under author supervision. AI agents
implemented and executed training and evaluation runs. AI agents
generated tables and figures used in this paper from logged
experiment results. AI agents checked formatting and citations. The
final manuscript text was written in full by the authors. All research
questions, experiment designs, scientific claims, and proofs originate
entirely from the authors.

\ifcameraready
\section{Acknowledgements}
This research was supported with Cloud TPUs from Google's TPU Research
Cloud (TRC).
\fi

\bibliography{references}

\ifappendsupplement
\clearpage
\setcounter{secnumdepth}{2}
\setcounter{section}{0}
\begin{center}
  {\Large\bfseries LILAC: Technical Supplement}
\end{center}
\raggedbottom
\makeatletter
\renewenvironment{table}[1][]{%
  \par\addvspace{0.8\baselineskip}%
  \noindent\begin{minipage}{\linewidth}%
  \def\@captype{table}%
}{%
  \end{minipage}%
  \par\addvspace{0.8\baselineskip}%
}
\renewenvironment{table*}[1][]{%
  \par\addvspace{0.8\baselineskip}%
  \noindent\begin{minipage}{\linewidth}%
  \def\@captype{table}%
}{%
  \end{minipage}%
  \par\addvspace{0.8\baselineskip}%
}
\makeatother

\begin{abstract}
  This supplement gives the complete quality, drift, ablation, downstream,
  and listening-test results behind the main paper. Metric directions are
  shown in each header; unavailable measurements are left blank.
\end{abstract}

\section{Reading the tables}

LibriTTS-R and LibriSpeech are reported separately. In the full comparison
tables, bold marks the best value within the sub-1\,kb/s or context band;
the high-rate DAC ladder is reference-only. Some compact table headers
abbreviate SCOREQ as SCQ. Ground-truth UTMOS is measured,
while its other metric entries are identity-signal ceilings. All codecs are
decoded at native rate and resampled only at corpus boundaries.

\section{Evaluation protocol details}
\label{sec:protocol}

Each codec is decoded once at native sample rate, freed from accelerator
memory, and scored from the cached waveform. Tables use full test splits
unless the caption gives a sample count.

\begin{table}[!htbp]
  \centering
  \footnotesize
  \begin{tabular}{@{}p{0.27\linewidth}p{0.66\linewidth}@{}}
    \toprule
    Metric & Definition \\
    \midrule
    UTMOS & SpeechMOS v1.2.0 non-reference MOS proxy
    (Saeki et al., 2022). \\
    dWER, quality tables & Whisper-large-v3 greedy WER between the
    reconstruction transcript and the original-audio transcript
    (Radford et al., 2023). The ground-truth row is therefore 0.000. \\
    dWER, drift tables & Whisper-large-v3 greedy WER against one fixed
    transcript of the original clip across every cycle. \\
    SCOREQ & Released reference-mode model (Ragano et al., 2024),
    applied with the matched original as the reference signal. \\
    PESQ & Wideband PESQ (Rix et al., 2001). \\
    STOI & Short-time objective intelligibility (Taal et al., 2010). \\
    SI-SNR & Scale-invariant signal-to-noise ratio
    (Le Roux et al., 2019). \\
    spk-sim & Cosine similarity of WavLM-SV x-vectors
    (\texttt{microsoft/wavlm-base-plus-sv}, VoxCeleb1-finetuned;
    Chen et al., 2022). \\
    \bottomrule
  \end{tabular}
  \caption{Metric implementations used throughout the supplement.}
  \label{tab:metric-implementations}
\end{table}

\begin{table}[!htbp]
  \centering
  \footnotesize
  \begin{tabular}{@{}lccc@{}}
    \toprule
    Evaluation & SCOREQ & UTMOS & dWER \\
    \midrule
    LibriSpeech, 600k & 0.011 & 0.021 & 0.002 \\
    LibriTTS-R, 600k & 0.010 & 0.022 & 0.010 \\
    \bottomrule
  \end{tabular}
  \caption{Observed flagship seed spread (standard deviation, $n=5$,
  SWA-10 condition).}
  \label{tab:seed-spread}
\end{table}

\clearpage
\onecolumn
\section{Full quality-comparison tables}
\label{sec:tables}

Each corpus has two aligned views: the first carries the metrics most useful
for a quick comparison; the second completes the signal and identity
measurements. The LILAC row label is boldfaced in both views. VCTK and H2 contain
only the sub-1\,kb/s panel.

\subsection{Codec specifications}

\begin{table}[!htbp]
  \centering
  \footnotesize
  \setlength{\tabcolsep}{4pt}
  \begin{tabular}{@{}p{0.21\textwidth}p{0.31\textwidth}rrrr@{}}
    \toprule
    Codec & Training corpus & Train h & Params (M) & Rate (Hz) & kb/s \\
    \midrule
    \textbf{LILAC} & HiFi-TTS 2 (full-band; 44.1\,kHz resampled to
    24\,kHz) &
    $\approx$31{,}720 & 58.5 & 9.375 & 0.75 \\
    \addlinespace
    FocalCodec (12.5\,Hz) & LibriTTS / train-clean-100$^\dagger$ &
    460$^\dagger$ & 142 (excl.\ WavLM) & 12.5 & 0.16 \\
    Mimi (nq=2) & undisclosed (large English) & --- & 82 & 12.5 & 0.275 \\
    FocalCodec (25\,Hz) & LibriTTS / train-clean-100$^\dagger$ &
    460$^\dagger$ & 142 (excl.\ WavLM) & 25 & 0.325 \\
    WavTokenizer (40\,Hz) & LibriTTS & $\sim$585 & $\sim$85 & 40 & 0.48 \\
    Mimi (nq=4) & undisclosed (large English) & --- & 82 & 12.5 & 0.55 \\
    FocalCodec (50\,Hz) & LibriTTS / train-clean-100$^\dagger$ &
    460$^\dagger$ & 142 (excl.\ WavLM) & 50 & 0.65 \\
    WavTokenizer (75\,Hz) & LibriTTS & $\sim$585 & $\sim$85 & 75 & 0.90 \\
    SNAC & undisclosed & --- & 19.8 & 47$^\S$ & 0.98 \\
    \addlinespace
    SpeechTokenizer (nq=2) & LibriSpeech & 960 & $\sim$108 & 50 & 1.0 \\
    BigCodec & LibriSpeech & 960 & $\sim$160 & 80 & 1.04 \\
    Mimi (nq=8) & undisclosed (large English) & --- & 82 & 12.5 & 1.1 \\
    DualCodec & Emilia & 100{,}000 & $\sim$95 & 12.5 & 1.23 \\
    EnCodec & DNS4 + CommonVoice + AudioSet + FSD50K + Jamendo &
    n/a$^\ddagger$ & $\sim$15 & 75 & 1.5 / 3 \\
    LFSC & MLS-en + CommonVoice & $\sim$28{,}700 & 112.7 & 21.5 & 1.89 \\
    SpeechTokenizer (nq=4) & LibriSpeech & 960 & $\sim$108 & 50 & 2.0 \\
    FunCodec & LibriTTS & $\sim$585 & 15.1 & 50 & 2 \\
    HiFi-Codec & LibriTTS + VCTK + AISHELL & $>$1{,}000 &
    $\sim$63 & 75 & 3 \\
    SpeechTokenizer (nq=8) & LibriSpeech & 960 & $\sim$108 & 50 & 4.0 \\
    \addlinespace
    DAC & DAPS + DNS4 + CommonVoice + VCTK + MUSDB + \ldots &
    n/a$^\ddagger$ & $\sim$74 & 75 & 3 / 6 / 12 / 24 \\
    \bottomrule
  \end{tabular}
  \caption{Released checkpoint specifications.}
  \label{tab:specs}
\end{table}

\begin{table}[!htbp]
  \centering
  \footnotesize
  \begin{minipage}[t]{0.48\textwidth}
    \centering
    \begin{tabular}{@{}lll@{}}
      \toprule
      Codec & LibriSpeech & LibriTTS-R \\
      \midrule
      \textbf{LILAC} & OOD & OOD \\
      FocalCodec (12.5\,Hz) & OOD & in-domain \\
      Mimi (nq=2) & undisclosed & undisclosed \\
      FocalCodec (25\,Hz) & OOD & in-domain \\
      WavTokenizer (40\,Hz) & OOD & in-domain \\
      Mimi (nq=4) & undisclosed & undisclosed \\
      FocalCodec (50\,Hz) & OOD & in-domain \\
      WavTokenizer (75\,Hz) & OOD & in-domain \\
      SNAC & undisclosed & undisclosed \\
      SpeechTokenizer (nq=2) & in-domain & OOD \\
      BigCodec & in-domain & OOD \\
      \bottomrule
    \end{tabular}
  \end{minipage}\hfill
  \begin{minipage}[t]{0.48\textwidth}
    \centering
    \begin{tabular}{@{}lll@{}}
      \toprule
      Codec & LibriSpeech & LibriTTS-R \\
      \midrule
      Mimi (nq=8) & undisclosed & undisclosed \\
      DualCodec & OOD & OOD \\
      EnCodec & OOD$^\ddagger$ & OOD \\
      LFSC & in-domain & OOD \\
      SpeechTokenizer (nq=4) & in-domain & OOD \\
      FunCodec & OOD & in-domain \\
      HiFi-Codec & OOD & in-domain \\
      SpeechTokenizer (nq=8) & in-domain & OOD \\
      DAC & OOD$^\ddagger$ & OOD \\
      \bottomrule
    \end{tabular}
  \end{minipage}
  \caption{Checkpoint-specific domain status.}
  \label{tab:spec-domains}
\end{table}

\noindent\footnotesize
\textit{Specification notes.}
$^\dagger$Released FocalCodec sources report both train-clean-100
(460\,h) and LibriTTS (960\,h). $^\ddagger$EnCodec and DAC do not publish
aggregate hours for their multi-corpus mixes; their LibriSpeech status is
approximate. FocalCodec parameter counts exclude its frozen WavLM front-end.
$^\S$SNAC uses 12/23/47\,Hz codebooks ($\sim$82 tokens/s total); 47\,Hz is
the finest grid.\normalsize

\newcommand{\LTSRRows}{%
  \MetricRow{Ground truth}{---}{---}
    {4.194 & 0.000$^\P$ & 4.64$^\P$ & 1.000$^\P$ & 0.000$^\P$}
    {$\infty^\P$ & 0.000$^\P$ & 0.000$^\P$ & 1.000$^\P$}
  \MetricBand{$<$1\,kb/s band}
  \MetricRow{FocalCodec (12.5\,Hz)}{in-domain}{0.16}
    {4.217 & 0.140 & 1.14 & 0.781 & 0.293}
    {$-$35.6 & 37.0 & 1.088 & 0.904}
  \MetricRow{Mimi (nq=2)}{undisclosed}{0.275}
    {2.537 & 0.199 & 1.22 & 0.780 & 0.734}
    {$-$16.9 & 41.0 & 0.946 & 0.815}
  \MetricRow{FocalCodec (25\,Hz)}{in-domain}{0.325}
    {4.124 & 0.085 & 1.22 & 0.836 & 0.247}
    {$-$34.4 & 32.4 & 0.953 & 0.930}
  \MetricRow{WavTokenizer (40\,Hz)}{in-domain}{0.48}
    {3.738 & 0.164 & 1.56 & 0.866 & 0.299}
    {$-$27.4 & 29.4 & 0.550 & 0.910}
  \MetricRow{Mimi (nq=4)}{undisclosed}{0.55}
    {3.294 & 0.125 & 1.66 & 0.869 & 0.498}
    {$-$1.1 & 32.6 & 0.743 & 0.907}
  \MetricRow{FocalCodec (50\,Hz)}{in-domain}{0.65}
    {4.076 & \textbf{0.058} & 1.33 & 0.868 & 0.216}
    {$-$32.9 & 28.9 & 0.875 & 0.950}
  \MetricRow{\textbf{LILAC (ours)}}{OOD}{0.75}
    {\textbf{4.238} & 0.086 & \textbf{2.60} & \textbf{0.944} & 0.252}
    {\textbf{+6.0} & 25.3 & 0.509 & 0.947}
  \MetricRow{WavTokenizer (75\,Hz)}{in-domain}{0.9}
    {4.182 & 0.079 & 2.41 & 0.923 & \textbf{0.164}}
    {+4.6 & \textbf{23.4} & \textbf{0.444} & 0.951}
  \MetricRow{SNAC}{undisclosed}{0.98}
    {3.951 & 0.067 & 2.24 & 0.918 & 0.192}
    {+2.3 & 23.5 & 0.468 & \textbf{0.953}}
  \MetricBand{$\ge$1\,kb/s context band}
  \MetricRow{SpeechTokenizer (nq=2)}{OOD}{1.0}
    {2.489 & 0.118 & 1.25 & 0.810 & 0.902}
    {$-$3.9 & 38.8 & 0.953 & 0.846}
  \MetricRow{BigCodec}{OOD}{1.04}
    {4.232 & 0.059 & 2.70 & 0.944 & 0.130}
    {+4.7 & 20.3 & 0.713 & 0.973}
  \MetricRow{Mimi (nq=8)}{undisclosed}{1.1}
    {3.823 & 0.079 & 2.30 & 0.918 & 0.282}
    {+4.4 & 26.2 & 0.626 & 0.952}
  \MetricRow{DualCodec}{OOD}{1.23}
    {4.252 & 0.049 & 2.81 & 0.950 & 0.139}
    {+7.5 & 19.3 & 0.419 & 0.979}
  \MetricRow{Encodec}{OOD}{1.5}
    {1.819 & 0.115 & 1.60 & 0.870 & 1.008}
    {+2.2 & 28.8 & 0.629 & 0.893}
  \MetricRow{LFSC}{OOD}{1.89}
    {\textbf{4.268} & \textbf{0.038} & 3.03 & \textbf{0.963} & 0.145}
    {+7.4 & 18.6 & 0.497 & \textbf{0.983}}
  \MetricRow{SpeechTokenizer (nq=4)}{OOD}{2.0}
    {3.722 & 0.076 & 1.94 & 0.899 & 0.322}
    {+1.6 & 27.9 & 0.786 & 0.929}
  \MetricRow{FunCodec}{in-domain}{2}
    {4.136 & 0.049 & 2.88 & 0.949 & 0.171}
    {+6.6 & 19.7 & 0.770 & 0.972}
  \MetricRow{Encodec}{OOD}{3}
    {2.753 & 0.062 & 2.12 & 0.918 & 0.571}
    {+4.8 & 21.7 & 0.530 & 0.944}
  \MetricRow{HiFi-Codec}{in-domain}{3}
    {4.134 & \textbf{0.038} & \textbf{3.21} & 0.955 & \textbf{0.114}}
    {\textbf{+7.7} & \textbf{17.5} & \textbf{0.365} & 0.980}
  \MetricRow{SpeechTokenizer (nq=8)}{OOD}{4.0}
    {4.081 & 0.050 & 2.67 & 0.935 & 0.169}
    {+4.7 & 21.9 & 0.691 & 0.969}
  \MetricBand{DAC high-rate reference}
  \MetricRow{DAC}{OOD$^\ddagger$}{3}
    {3.500 & 0.060 & 2.61 & 0.930 & 0.333}
    {$-$3.3 & 21.8 & 0.434 & 0.952}
  \MetricRow{DAC}{OOD$^\ddagger$}{6}
    {3.998 & 0.038 & 3.65 & 0.964 & 0.142}
    {$-$3.7 & 15.4 & 0.337 & 0.980}
  \MetricRow{DAC}{OOD$^\ddagger$}{12}
    {4.150 & 0.027 & 4.28 & 0.984 & 0.070}
    {$-$3.8 & 9.8 & 0.244 & 0.992}
  \MetricRow{DAC}{OOD$^\ddagger$}{24}
    {4.184 & 0.022 & 4.48 & 0.994 & 0.051}
    {$-$3.8 & 5.5 & 0.164 & 0.996}
}

\begin{table}[!htbp]
  \centering
  \small
  \renewcommand{\arraystretch}{1.60}
  \setlength{\tabcolsep}{6pt}
  \begingroup
  \def\MetricRow#1#2#3#4#5{#1 & #2 & #3 & #4 \\}
  \def\MetricBand#1{\midrule\multicolumn{8}{l}{\textit{#1}} \\}
  \begin{tabular}{@{}lllrrrrr@{}}
    \toprule
    Codec & Domain & kb/s & UTMOS$\uparrow$ & dWER$\downarrow$ &
    PESQ$\uparrow$ & STOI$\uparrow$ & SCOREQ$\downarrow$ \\
    \midrule
    \LTSRRows
    \bottomrule
  \end{tabular}
  \endgroup
  \caption{LibriTTS-R test: priority metrics for all evaluated codecs
  (n=4837). Bold marks the best value in each rate band.}
  \label{tab:ltsr-full}
\end{table}

\begin{table}[!htbp]
  \centering
  \small
  \renewcommand{\arraystretch}{1.60}
  \setlength{\tabcolsep}{9pt}
  \begingroup
  \def\MetricRow#1#2#3#4#5{#1 & #3 & #5 \\}
  \def\MetricBand#1{\midrule\multicolumn{6}{l}{\textit{#1}} \\}
  \begin{tabular}{@{}llrrrr@{}}
    \toprule
    Codec & kb/s & SI-SNR$\uparrow$ (dB) & MCD$\downarrow$ &
    mel-dist$\downarrow$ & spk-sim$\uparrow$ \\
    \midrule
    \LTSRRows
    \bottomrule
  \end{tabular}
  \endgroup
  \caption{LibriTTS-R test: signal and speaker-identity metrics, in the
  same row order as the priority view.}
  \label{tab:ltsr-signal}
\end{table}

\newcommand{\LSRows}{%
  \MetricRow{Ground truth}{---}{---}
    {4.072 & 0.000$^\P$ & 4.64$^\P$ & 1.000$^\P$ & 0.000$^\P$}
    {$\infty^\P$ & 0.000$^\P$ & 0.000$^\P$ & 1.000$^\P$}
  \MetricBand{$<$1\,kb/s band}
  \MetricRow{FocalCodec (12.5\,Hz)}{OOD}{0.16}
    {\textbf{4.226} & 0.131 & 1.19 & 0.780 & 0.317}
    {$-$37.4 & 35.9 & 0.784 & 0.924}
  \MetricRow{Mimi (nq=2)}{undisclosed}{0.275}
    {2.532 & 0.185 & 1.26 & 0.769 & 0.705}
    {$-$24.2 & 38.8 & 0.970 & 0.851}
  \MetricRow{FocalCodec (25\,Hz)}{OOD}{0.325}
    {4.145 & 0.079 & 1.29 & 0.830 & 0.275}
    {$-$36.1 & 31.4 & 0.693 & 0.953}
  \MetricRow{WavTokenizer (40\,Hz)}{OOD}{0.48}
    {3.577 & 0.194 & 1.61 & 0.849 & 0.351}
    {$-$30.4 & 28.5 & 0.626 & 0.925}
  \MetricRow{Mimi (nq=4)}{undisclosed}{0.55}
    {3.133 & 0.128 & 1.66 & 0.853 & 0.514}
    {$-$5.4 & 30.7 & 0.756 & 0.932}
  \MetricRow{FocalCodec (50\,Hz)}{OOD}{0.65}
    {4.059 & \textbf{0.067} & 1.39 & 0.859 & \textbf{0.242}}
    {$-$34.5 & 28.2 & 0.634 & \textbf{0.968}}
  \MetricRow{\textbf{LILAC (ours)}}{OOD}{0.75}
    {4.141 & 0.101 & \textbf{2.60} & \textbf{0.935} & 0.299}
    {\textbf{+2.2} & 24.6 & \textbf{0.499} & 0.956}
  \MetricRow{WavTokenizer (75\,Hz)}{OOD}{0.9}
    {3.783 & 0.087 & 2.11 & 0.896 & 0.252}
    {$-$1.0 & \textbf{23.5} & 0.527 & 0.952}
  \MetricRow{SNAC}{undisclosed}{0.98}
    {3.276 & 0.090 & 1.97 & 0.889 & 0.453}
    {$-$3.1 & 25.0 & 0.512 & 0.946}
  \MetricBand{$\ge$1\,kb/s context band}
  \MetricRow{SpeechTokenizer (nq=2)}{in-domain}{1.0}
    {2.193 & 0.134 & 1.27 & 0.782 & 0.903}
    {$-$7.8 & 37.5 & 0.722 & 0.894}
  \MetricRow{BigCodec}{in-domain}{1.04}
    {4.075 & 0.068 & 2.62 & 0.932 & \textbf{0.164}}
    {+0.5 & 20.0 & 0.504 & 0.978}
  \MetricRow{Mimi (nq=8)}{undisclosed}{1.1}
    {3.665 & 0.085 & 2.28 & 0.905 & 0.324}
    {+1.5 & 25.4 & 0.645 & 0.968}
  \MetricRow{DualCodec}{OOD}{1.23}
    {4.086 & 0.062 & 2.70 & 0.937 & 0.189}
    {\textbf{+4.3} & 18.7 & \textbf{0.416} & 0.983}
  \MetricRow{Encodec}{OOD}{1.5}
    {1.622 & 0.125 & 1.60 & 0.848 & 0.998}
    {$-$0.6 & 27.2 & 0.777 & 0.931}
  \MetricRow{LFSC}{in-domain}{1.89}
    {\textbf{4.129} & 0.055 & \textbf{3.07} & \textbf{0.955} & 0.198}
    {+4.2 & \textbf{18.5} & 0.565 & \textbf{0.987}}
  \MetricRow{SpeechTokenizer (nq=4)}{in-domain}{2.0}
    {3.457 & 0.099 & 1.91 & 0.880 & 0.381}
    {$-$1.8 & 27.1 & 0.556 & 0.954}
  \MetricRow{FunCodec}{OOD}{2}
    {3.876 & 0.059 & 2.60 & 0.929 & 0.225}
    {+0.8 & 19.6 & 0.565 & 0.977}
  \MetricRow{Encodec}{OOD}{3}
    {2.418 & 0.082 & 2.14 & 0.902 & 0.635}
    {+2.3 & 21.1 & 0.677 & 0.967}
  \MetricRow{HiFi-Codec}{OOD}{3}
    {3.727 & \textbf{0.053} & 2.65 & 0.932 & 0.273}
    {+2.8 & 19.8 & 0.493 & 0.973}
  \MetricRow{SpeechTokenizer (nq=8)}{in-domain}{4.0}
    {3.835 & 0.062 & 2.57 & 0.920 & 0.221}
    {+1.2 & 21.7 & 0.482 & 0.979}
  \MetricBand{DAC high-rate reference}
  \MetricRow{DAC}{OOD$^\ddagger$}{3}
    {2.985 & 0.079 & 2.44 & 0.910 & 0.469}
    {$-$8.9 & 22.8 & 0.442 & 0.958}
  \MetricRow{DAC}{OOD$^\ddagger$}{6}
    {3.670 & 0.054 & 3.50 & 0.954 & 0.212}
    {$-$9.5 & 16.1 & 0.341 & 0.984}
  \MetricRow{DAC}{OOD$^\ddagger$}{12}
    {3.961 & 0.037 & 4.22 & 0.982 & 0.093}
    {$-$9.5 & 10.2 & 0.245 & 0.995}
  \MetricRow{DAC}{OOD$^\ddagger$}{24}
    {4.041 & 0.028 & 4.46 & 0.995 & 0.060}
    {$-$9.5 & 5.6 & 0.161 & 0.998}
}

\begin{table}[!htbp]
  \centering
  \small
  \renewcommand{\arraystretch}{1.60}
  \setlength{\tabcolsep}{6pt}
  \begingroup
  \def\MetricRow#1#2#3#4#5{#1 & #2 & #3 & #4 \\}
  \def\MetricBand#1{\midrule\multicolumn{8}{l}{\textit{#1}} \\}
  \begin{tabular}{@{}lllrrrrr@{}}
    \toprule
    Codec & Domain & kb/s & UTMOS$\uparrow$ & dWER$\downarrow$ &
    PESQ$\uparrow$ & STOI$\uparrow$ & SCOREQ$\downarrow$ \\
    \midrule
    \LSRows
    \bottomrule
  \end{tabular}
  \endgroup
  \caption{LibriSpeech test-clean: priority metrics for all evaluated
  codecs (n=2620). Bold marks the best value in each rate band.}
  \label{tab:ls-full}
\end{table}

\begin{table}[!htbp]
  \centering
  \small
  \renewcommand{\arraystretch}{1.60}
  \setlength{\tabcolsep}{9pt}
  \begingroup
  \def\MetricRow#1#2#3#4#5{#1 & #3 & #5 \\}
  \def\MetricBand#1{\midrule\multicolumn{6}{l}{\textit{#1}} \\}
  \begin{tabular}{@{}llrrrr@{}}
    \toprule
    Codec & kb/s & SI-SNR$\uparrow$ (dB) & MCD$\downarrow$ &
    mel-dist$\downarrow$ & spk-sim$\uparrow$ \\
    \midrule
    \LSRows
    \bottomrule
  \end{tabular}
  \endgroup
  \caption{LibriSpeech test-clean: signal and speaker-identity metrics,
  in the same row order as the priority view.}
  \label{tab:ls-signal}
\end{table}

\newcommand{\VCTKRows}{%
  \MetricRow{Ground truth}{---}{---}
    {4.060 & 0.000$^\P$ & 4.64$^\P$ & 1.000$^\P$ & 0.000$^\P$}
    {$\infty^\P$ & 0.000$^\P$ & 0.000$^\P$ & 1.000$^\P$}
  \MetricBand{$<$1\,kb/s band}
  \MetricRow{FocalCodec (12.5\,Hz)}{OOD}{0.16}
    {\textbf{4.143} & 0.083 & 1.27 & 0.746 & 0.350}
    {$-$33.8 & 31.9 & 0.901 & 0.853}
  \MetricRow{Mimi (nq=2)}{undisclosed}{0.275}
    {2.858 & 0.094 & 1.40 & 0.722 & 0.694}
    {$-$25.0 & 33.1 & 0.810 & 0.759}
  \MetricRow{FocalCodec (25\,Hz)}{OOD}{0.325}
    {4.083 & 0.044 & 1.42 & 0.793 & 0.289}
    {$-$32.2 & 27.3 & 0.815 & 0.920}
  \MetricRow{WavTokenizer (40\,Hz)}{OOD}{0.48}
    {3.769 & 0.118 & 1.63 & 0.806 & 0.337}
    {$-$27.2 & 23.5 & 0.473 & 0.876}
  \MetricRow{Mimi (nq=4)}{undisclosed}{0.55}
    {3.345 & 0.057 & 1.73 & 0.787 & 0.547}
    {$-$3.8 & 26.1 & 0.672 & 0.837}
  \MetricRow{FocalCodec (50\,Hz)}{OOD}{0.65}
    {4.024 & 0.032 & 1.56 & 0.818 & 0.262}
    {$-$31.4 & 24.5 & 0.760 & \textbf{0.939}}
  \MetricRow{\textbf{LILAC (ours)}}{OOD}{0.75}
    {4.104 & 0.041 & \textbf{2.59} & \textbf{0.867} & 0.312}
    {\textbf{+5.3} & 21.0 & 0.468 & 0.921}
  \MetricRow{WavTokenizer (75\,Hz)}{OOD}{0.90}
    {3.587 & 0.047 & 1.82 & 0.835 & 0.270}
    {$-$5.0 & 20.3 & \textbf{0.426} & 0.909}
  \MetricRow{SNAC}{undisclosed}{0.98}
    {3.810 & \textbf{0.029} & 2.29 & 0.849 & \textbf{0.242}}
    {$-$1.3 & \textbf{18.8} & 0.432 & 0.917}
}

\begin{table}[!htbp]
  \centering
  \footnotesize
  \textbf{A. Priority metrics}\par\smallskip
  \setlength{\tabcolsep}{6pt}
  \begingroup
  \def\MetricRow#1#2#3#4#5{#1 & #2 & #3 & #4 \\}
  \def\MetricBand#1{\midrule\multicolumn{8}{l}{\textit{#1}} \\}
  \begin{tabular}{@{}lllrrrrr@{}}
    \toprule
    Codec & Domain & kb/s & UTMOS$\uparrow$ & dWER$\downarrow$ &
    PESQ$\uparrow$ & STOI$\uparrow$ & SCOREQ$\downarrow$ \\
    \midrule
    \VCTKRows
    \bottomrule
  \end{tabular}
  \endgroup
  \par\smallskip
  \footnotesize Nominal rates differ across rows.
  \par\vspace{1.2\baselineskip}
  \textbf{B. Signal and speaker-identity metrics}\par\smallskip
  \setlength{\tabcolsep}{9pt}
  \begingroup
  \def\MetricRow#1#2#3#4#5{#1 & #3 & #5 \\}
  \def\MetricBand#1{\midrule\multicolumn{6}{l}{\textit{#1}} \\}
  \begin{tabular}{@{}llrrrr@{}}
    \toprule
    Codec & kb/s & SI-SNR$\uparrow$ (dB) & MCD$\downarrow$ &
    mel-dist$\downarrow$ & spk-sim$\uparrow$ \\
    \midrule
    \VCTKRows
    \bottomrule
  \end{tabular}
  \endgroup
  \caption{VCTK test, sub-1\,kb/s panel (n=3094).}
  \label{tab:vctk-full}
  \label{tab:vctk-signal}
\end{table}

\newcommand{\HTwoRows}{%
  \MetricBand{$<$1\,kb/s external baselines}
  \MetricRow{FocalCodec (12.5\,Hz)}{undetermined}{0.16}
    {\textbf{4.162} & 0.126 & 1.20 & 0.781 & 0.290}
    {$-$36.2 & 35.3 & 1.020 & 0.944}
  \MetricRow{Mimi (nq=2)}{undetermined}{0.275}
    {2.423 & 0.207 & 1.24 & 0.766 & 0.684}
    {$-$20.9 & 38.2 & 0.890 & 0.874}
  \MetricRow{FocalCodec (25\,Hz)}{undetermined}{0.325}
    {4.054 & 0.074 & 1.31 & 0.829 & 0.242}
    {$-$35.4 & 31.1 & 0.925 & 0.968}
  \MetricRow{WavTokenizer (40\,Hz)}{undetermined}{0.48}
    {3.557 & 0.181 & 1.61 & 0.843 & 0.274}
    {$-$30.6 & 28.1 & 0.539 & 0.939}
  \MetricRow{Mimi (nq=4)}{undetermined}{0.55}
    {3.024 & 0.134 & 1.64 & 0.850 & 0.469}
    {$-$2.9 & 30.4 & 0.724 & 0.944}
  \MetricRow{FocalCodec (50\,Hz)}{undetermined}{0.65}
    {3.978 & \textbf{0.067} & 1.42 & 0.856 & 0.211}
    {$-$34.0 & 28.1 & 0.860 & \textbf{0.977}}
  \MetricRow{WavTokenizer (75\,Hz)}{undetermined}{0.90}
    {3.642 & 0.093 & \textbf{2.01} & \textbf{0.889} & \textbf{0.190}}
    {\textbf{$-$0.6} & \textbf{23.3} & \textbf{0.467} & 0.964}
  \MetricRow{SNAC}{undetermined}{0.98}
    {3.429 & 0.079 & \textbf{2.01} & \textbf{0.889} & 0.259}
    {$-$1.9 & 23.7 & 0.488 & 0.964}
}

\begin{table}[!htbp]
  \centering
  \footnotesize
  \setlength{\tabcolsep}{6pt}
  \begingroup
  \def\MetricRow#1#2#3#4#5{#1 & #2 & #3 & #4 \\}
  \def\MetricBand#1{\midrule\multicolumn{8}{l}{\textit{#1}} \\}
  \begin{tabular}{@{}lllrrrrr@{}}
    \toprule
    Codec & Domain & kb/s & UTMOS$\uparrow$ & dWER$\downarrow$ &
    PESQ$\uparrow$ & STOI$\uparrow$ & SCOREQ$\downarrow$ \\
    \HTwoRows
    \bottomrule
  \end{tabular}
  \endgroup
  \caption{H2 evaluation panel: priority metrics for external baselines
  (n=585).}
  \label{tab:h2-full}
\end{table}

\begin{table}[!htbp]
  \centering
  \footnotesize
  \setlength{\tabcolsep}{9pt}
  \begingroup
  \def\MetricRow#1#2#3#4#5{#1 & #3 & #5 \\}
  \def\MetricBand#1{\midrule\multicolumn{6}{l}{\textit{#1}} \\}
  \begin{tabular}{@{}llrrrr@{}}
    \toprule
    Codec & kb/s & SI-SNR$\uparrow$ (dB) & MCD$\downarrow$ &
    mel-dist$\downarrow$ & spk-sim$\uparrow$ \\
    \HTwoRows
    \bottomrule
  \end{tabular}
  \endgroup
  \caption{H2 evaluation panel: signal and speaker-identity metrics for
  the same baselines as Table~\ref{tab:h2-full}.}
  \label{tab:h2-signal}
\end{table}

\begin{table}[!htbp]
  \centering
  \footnotesize
  \setlength{\tabcolsep}{5pt}
  \begin{tabular}{lrrrrrrrrr}
    \toprule
    & UTMOS & dWER & PESQ & STOI & SI-SNR & MCD & mel-dist & spk-sim & SCOREQ \\
    \midrule
    LILAC (ours) & 4.087 & 0.096 & 2.48 & 0.929 & +2.5 & 25.0 & 0.527
    & 0.962 & 0.267 \\
    \bottomrule
  \end{tabular}
  \par\smallskip
  \footnotesize This row is descriptive and excluded from comparisons with
  Table~\ref{tab:h2-full}; external-baseline reader overlap is unknown, and
  no ground-truth UTMOS was measured.
  \caption{LILAC on the H2 evaluation panel (n=585), shown separately
  because this panel is in-domain for LILAC: its readers are held out
  from training by ID, but come from the training corpus, unlike the
  out-of-domain evaluation sets.}
  \label{tab:h2-lilac}
\end{table}

\subsection{SWA-10 @ 400k comparator}
\label{sec:swa400}

Table~\ref{tab:swa400comparator} reports the earlier SWA-10 comparator
(checkpoints 391k--400k). The flagship averages checkpoints 889k--898k
from the same run and improves UTMOS, PESQ, and SCOREQ on all three
corpora, with a dWER cost on LibriSpeech.

\begin{table}[!htbp]
  \centering
  \footnotesize
  \setlength{\tabcolsep}{1.7pt}
  \begin{tabular}{lrrrrr}
    \toprule
    Corpus & UTMOS & dWER & PESQ & SI-SNR & SCQ \\
    \midrule
    LibriSpeech & 4.108 & 0.102 & 2.58 & $+2.0$ & 0.306 \\
    LibriTTS-R  & 4.209 & 0.086 & 2.58 & $+5.9$ & 0.264 \\
    VCTK        & 4.069 & \textbf{0.043} & 2.59 & $+5.1$ & 0.316 \\
    \bottomrule
  \end{tabular}
  \caption{LILAC SWA-10 @ 400k on the three evaluation corpora.}
  \label{tab:swa400comparator}
\end{table}

\subsection{Full component ablation}
\label{sec:ablation}

The three arms isolate the invertible anti-imaging stem and the
target-preserving attenuation augmentation at a matched 300k steps.

\begin{table}[!htbp]
  \centering
  \footnotesize
  \setlength{\tabcolsep}{2.5pt}
  \begin{tabular}{lcccccc}
    \toprule
    & \multicolumn{2}{c}{SCOREQ $\downarrow$} &
    \multicolumn{2}{c}{UTMOS $\uparrow$} & \multicolumn{2}{c}{dWER
    $\downarrow$} \\
    \cmidrule(lr){2-3}\cmidrule(lr){4-5}\cmidrule(lr){6-7}
    Configuration & LS & LT-R & LS & LT-R & LS & LT-R \\
    \midrule
    neither                & 0.531 & 0.572 & 3.66 & 3.74 & 0.124 & 0.121 \\
    stem only              & \textbf{0.346} & 0.604 & 3.98 & 3.42 &
    0.115 & 0.252 \\
    stem + aug.\ (full)    & 0.352 & \textbf{0.327} & \textbf{4.00} &
    \textbf{4.10} & \textbf{0.109} & \textbf{0.098} \\
    \bottomrule
  \end{tabular}
  \caption{Component ablation at 300k steps on LibriSpeech (LS) and
  LibriTTS-R (LT-R).}
  \label{tab:ablation}
\end{table}

\subsection{Cost-of-exactness probe details}
\label{sec:exactness}

These smaller, earlier in-domain probes isolate the cost of exactness:
\begin{itemize}
  \item Untying the decoder gives an untied-minus-tied LibriSpeech UTMOS
    gap of $-0.004$ at 25k and $+0.185$ at 150k, with approximately
    $2\times$ the transform parameters.
  \item A 13.8M-parameter free-form decoder reaches UTMOS 3.56 at 300k
    (3.46 at 150k); the invertible decoder reaches 3.75 on the same codes.
  \item A fill-network capacity sweep finds no improvement over the
    reported 0.75\,kb/s model.
\end{itemize}

\section{Full re-encode-drift table}
\label{sec:drift}

Five codecs are evaluated at cycles 1, 2, 5, 10, 50, and 100 on the same
100 reader-balanced LibriSpeech test-clean clips. Code-match is the fraction
of tokens equal to the first encoding or the previous cycle. Conclusions
are within-codec because rates range from 0.65 to approximately 24\,kb/s.
LILAC uses a low-quality probe checkpoint here; its flat trajectory tests
the structural fixed-point property, not flagship quality.

\newcommand{\DriftRows}{%
  \DriftBand{\textbf{LILAC} (0.75\,kb/s; probe checkpoint)}
  \DriftRow{LILAC}{1}{1.0000}{1.0000}{1.455 & 0.806 & $-$25.73 & 1.299}
  \DriftRow{LILAC}{2}{1.0000}{1.0000}{1.455 & 0.806 & $-$25.73 & 1.299}
  \DriftRow{LILAC}{5}{1.0000}{1.0000}{1.455 & 0.806 & $-$25.73 & 1.299}
  \DriftRow{LILAC}{10}{1.0000}{1.0000}{1.455 & 0.806 & $-$25.73 & 1.299}
  \DriftRow{LILAC}{50}{1.0000}{1.0000}{1.455 & 0.806 & $-$25.73 & 1.299}
  \DriftRow{LILAC}{100}{1.0000}{1.0000}{1.455 & 0.806 & $-$25.73 & 1.299}
  \DriftBand{Encodec (1.5\,kb/s)}
  \DriftRow{Encodec}{1}{0.8047}{0.8047}{1.617 & 0.847 & $-$0.12 & 1.615}
  \DriftRow{Encodec}{2}{0.7667}{0.8983}{1.566 & 0.836 & $-$0.45 & 1.558}
  \DriftRow{Encodec}{5}{0.7352}{0.9115}{1.527 & 0.830 & $-$0.61 & 1.526}
  \DriftRow{Encodec}{10}{0.7235}{0.9587}{1.508 & 0.828 & $-$0.64 & 1.515}
  \DriftRow{Encodec}{50}{0.7144}{0.9783}{1.493 & 0.827 & $-$0.66 & 1.512}
  \DriftRow{Encodec}{100}{0.7141}{0.9994}{1.492 & 0.827 & $-$0.66 & 1.511}
  \DriftBand{DAC ($\sim$24\,kb/s; high-rate reference)}
  \DriftRow{DAC}{1}{0.0173}{0.0173}{4.460 & 0.995 & $-$8.48 & 4.058}
  \DriftRow{DAC}{2}{0.0108}{0.0172}{4.362 & 0.989 & $-$17.11 & 4.027}
  \DriftRow{DAC}{5}{0.0085}{0.0093}{3.815 & 0.965 & $-$20.13 & 3.863}
  \DriftRow{DAC}{10}{0.0078}{0.0094}{2.516 & 0.909 & $-$22.15 & 3.408}
  \DriftRow{DAC}{50}{0.0056}{0.0063}{1.046 & 0.452 & $-$32.07 & 1.253}
  \DriftRow{DAC}{100}{0.0049}{0.0071}{1.033 & 0.211 & $-$41.02 & 1.260}
  \DriftBand{Mimi ($\sim$1.1\,kb/s; semantic-distilled)}
  \DriftRow{Mimi}{1}{0.3289}{0.3289}{2.298 & 0.903 & 2.21 & 3.695}
  \DriftRow{Mimi}{2}{0.2209}{0.3892}{1.979 & 0.872 & 0.88 & 3.487}
  \DriftRow{Mimi}{5}{0.1297}{0.2544}{1.532 & 0.796 & $-$1.08 & 2.826}
  \DriftRow{Mimi}{10}{0.0807}{0.2846}{1.272 & 0.681 & $-$2.89 & 1.851}
  \DriftRow{Mimi}{50}{0.0648}{0.4330}{1.160 & 0.557 & $-$4.66 & 1.475}
  \DriftRow{Mimi}{100}{0.0647}{0.9900}{1.161 & 0.557 & $-$4.67 & 1.476}
  \DriftBand{FocalCodec (0.65\,kb/s)}
  \DriftRow{FocalCodec}{1}{0.1087}{0.1087}{1.413 & 0.859 & $-$34.49 & 4.059}
  \DriftRow{FocalCodec}{2}{0.0410}{0.2074}{1.223 & 0.812 & $-$36.46 & 4.057}
  \DriftRow{FocalCodec}{5}{0.0096}{0.0654}{1.105 & 0.719 & $-$38.77 & 3.823}
  \DriftRow{FocalCodec}{10}{0.0037}{0.0453}{1.062 & 0.621 & $-$38.79 & 3.482}
  \DriftRow{FocalCodec}{50}{0.0004}{0.0023}{1.041 & 0.365 & $-$46.86 & 2.796}
  \DriftRow{FocalCodec}{100}{0.0005}{0.0040}{1.043 & 0.277 & $-$46.69 & 2.677}
}

\begin{table}[!htbp]
  \centering
  \footnotesize
  \setlength{\tabcolsep}{12pt}
  \begingroup
  \def\DriftRow#1#2#3#4#5{#1 & #2 & #3 & #4 \\}
  \def\DriftBand#1{\midrule\multicolumn{4}{l}{\textit{#1}} \\}
  \begin{tabular}{@{}llrr@{}}
    \toprule
    Codec & Cycle & Match first $\uparrow$ & Match previous $\uparrow$ \\
    \DriftRows
    \bottomrule
  \end{tabular}
  \endgroup
  \caption{Token agreement under repeated re-encoding. LILAC remains at
  1.0000 through cycle 100.}
  \label{tab:drift-lilac}
  \label{tab:drift-baselines}
\end{table}

\begin{table}[!htbp]
  \centering
  \footnotesize
  \setlength{\tabcolsep}{10pt}
  \begin{tabular}{@{}lrrrr@{}}
    \toprule
    Codec & kb/s & dWER (cycle 1) & dWER (cycle 10) & dWER (cycle 100) \\
    \midrule
    \textbf{LILAC (probe)} & 0.75 & 0.4680 &
    0.4680 & 0.4680 \\
    Encodec & 1.5 & 0.1146 & 0.1383 & \textbf{0.1429} \\
    Mimi & 1.1 & 0.0988 & 0.7461 & 1.0124 \\
    FocalCodec & 0.65 & 0.0718 & 0.4893 & 1.3352 \\
    DAC & $\sim$24 & \textbf{0.0244} & \textbf{0.0866} & 1.4501 \\
    \bottomrule
  \end{tabular}
  \caption{dWER at representative re-encoding depths. The reference
  transcript is fixed across cycles.}
  \label{tab:drift-dwer}
\end{table}

\begin{table}[!htbp]
  \centering
  \footnotesize
  \setlength{\tabcolsep}{12pt}
  \begingroup
  \def\DriftRow#1#2#3#4#5{#1 & #2 & #5 \\}
  \def\DriftBand#1{\midrule\multicolumn{6}{l}{\textit{#1}} \\}
  \begin{tabular}{@{}llrrrr@{}}
    \toprule
    Codec & Cycle & PESQ$\uparrow$ & STOI$\uparrow$ &
    SI-SNR$\uparrow$ (dB) & UTMOS$\uparrow$ \\
    \DriftRows
    \bottomrule
  \end{tabular}
  \endgroup
  \caption{Quality under repeated re-encoding. LILAC is constant; the
  baseline trajectories degrade or settle onto different code streams.}
  \label{tab:drift-quality}
\end{table}

\noindent\footnotesize
The drift panel contains 100 clips and ends at cycle 100. SI-SNR is retained
as a secondary metric because vocoder alignment can move it differently
from perceptual measures.\normalsize

\subsection{Broader single-pass panel (12 configurations)}

The broader panel uses the same seeded 50-clip LibriTTS-R subset for all
12 baseline configurations (11 codec families; EnCodec appears at two
bitrates). \texttt{agree@}$N$ is agreement with the first encoding;
$\Delta$UTMOS@100 is the change from cycle 1. Every baseline rewrites at
least 15\% of its tokens on the first pass. SpeechTokenizer illustrates
the distinction between token and waveform stability: UTMOS changes by
$+0.00$ while agreement falls to 0.573.

\par\smallskip
\begingroup
  \centering
  \footnotesize
  \setlength{\tabcolsep}{2.5pt}
  \begin{tabular}{lccr}
    \toprule
    Codec & agree@1 $\uparrow$ & agree@100 $\uparrow$ & $\Delta$UTMOS@100 \\
    \midrule
    BigCodec         & 0.847 & 0.781 & $-0.01$ \\
    EnCodec (1.5\,kb/s) & 0.844 & 0.761 & $-0.12$ \\
    SNAC             & 0.758 & 0.539 & $-0.13$ \\
    SpeechTokenizer  & 0.726 & 0.573 & $+0.00$ \\
    EnCodec (3.0\,kb/s) & 0.647 & 0.407 & $-0.77$ \\
    HiFi-Codec       & 0.631 & 0.242 & $-1.03$ \\
    FunCodec         & 0.521 & 0.209 & $-2.53$ \\
    LFSC             & 0.416 & 0.047 & $-1.82$ \\
    Mimi             & 0.396 & 0.104 & $-2.07$ \\
    WavTokenizer     & 0.392 & 0.007 & $-1.55$ \\
    DualCodec        & 0.198 & 0.002 & $-2.45$ \\
    FocalCodec       & 0.109 & 0.000 & $-1.18$ \\
    \bottomrule
  \end{tabular}
  \par
\endgroup

\section{Additional evaluation panels}
\label{sec:additional}

This section reports the fixed-rate design grid and downstream transfer
measurements.

\subsection{Fixed-rate allocation grid}
\label{sec:rategrid}

Every quantized arm uses 0.75\,kb/s and the same training recipe while
varying frame rate, coordinate count, precision, or quantization; the
no-quantization diagnostic transmits continuous latents and has no fixed
bitrate. Arm rows are matched
within each labeled block; blocks use the stated checkpoints and are not
compared across training horizons. The flagship rows in the first block are
its shipping read (SWA-10 @898k, main Table~2), included as a reference
point rather than a matched-horizon cell. SWA-$k$ averages the final $k$
checkpoints at 1k-step intervals.

\begin{table*}[t]
  \centering
  \footnotesize
  \setlength{\tabcolsep}{3pt}
  \begin{tabular}{@{}p{0.30\textwidth}lrrrrrr@{}}
    \toprule
    Configuration & Corpus & kb/s & Hz & bits/coord & UTMOS & dWER & SCOREQ \\
    \midrule
    \multicolumn{8}{l}{\textit{Coordinates vs.\ frame rate (arms: SWA-10
    @150k; flagship reference: SWA-10 @898k)}} \\
    20 coordinates (flagship @898k, main Table~2) & LS  & 0.75 & 9.375 & 4 &
    4.141 & 0.101 & 0.299 \\
    20 coordinates (flagship @898k, main Table~2) & LT-R & 0.75 & 9.375 & 4 &
    4.238 & 0.086 & 0.252 \\
    10 coordinates & LS  & 0.75 & 18.75 & 4 & 3.589 & 0.124 & 0.410 \\
    10 coordinates & LT-R & 0.75 & 18.75 & 4 & 3.711 & 0.101 & 0.368 \\
    5 coordinates & LS  & 0.75 & 37.5 & 4 & 3.824 & 0.127 & 0.419 \\
    5 coordinates & LT-R & 0.75 & 37.5 & 4 & 3.961 & 0.112 & 0.406 \\
    \midrule
    \multicolumn{8}{l}{\textit{Precision, mature read (SWA-3 @150k,
    matched $k$)}} \\
    10 coordinates $\times$ 4 bit (comparator) & LS  & 0.75 & 18.75 &
    4 & 3.555 & 0.119 & 0.427 \\
    10 coordinates $\times$ 4 bit (comparator) & LT-R & 0.75 & 18.75
    & 4 & 3.683 & 0.106 & 0.383 \\
    20 coordinates $\times$ 2 bit & LS  & 0.75 & 18.75 & 2 & 3.856 &
    0.127 & 0.392 \\
    20 coordinates $\times$ 2 bit & LT-R & 0.75 & 18.75 & 2 & 3.968 &
    0.118 & 0.387 \\
    \midrule
    \multicolumn{8}{l}{\textit{Precision floor screen (raw @50k)}} \\
    10 coordinates $\times$ 4 bit (comparator) & LS  & 0.75 & 18.75 &
    4 & 2.990 & 0.149 & 0.657 \\
    10 coordinates $\times$ 4 bit (comparator) & LT-R & 0.75 & 18.75
    & 4 & 3.111 & 0.143 & 0.649 \\
    20 coordinates $\times$ 2 bit (comparator) & LS  & 0.75 & 18.75 &
    2 & 3.395 & 0.154 & 0.643 \\
    20 coordinates $\times$ 2 bit (comparator) & LT-R & 0.75 & 18.75
    & 2 & 3.508 & 0.153 & 0.674 \\
    40 coordinates $\times$ 1 bit & LS  & 0.75 & 18.75 & 1 & 2.799 &
    0.464 & 0.876 \\
    40 coordinates $\times$ 1 bit & LT-R & 0.75 & 18.75 & 1 & 2.984 &
    0.480 & 0.878 \\
    \midrule
    \multicolumn{8}{l}{\textit{Quantization (raw, matched step 150k)}} \\
    Flagship control (re-eval spread) & LS & 0.75 & 9.375 & 4 &
    3.868--3.869 & 0.120--0.126 & 0.426--0.432 \\
    Flagship control (re-eval spread) & LT-R & 0.75 & 9.375 & 4 &
    4.009--4.019 & 0.103--0.105 & 0.400--0.407 \\
    No quantization (continuous latents) & LS & --- & 9.375 & --- &
    3.604 & 0.120 & 0.535 \\
    No quantization (continuous latents) & LT-R & --- & 9.375 & ---
    & 3.789 & 0.117 & 0.510 \\
    \bottomrule
  \end{tabular}
  \caption{Fixed-rate allocation grid on LibriSpeech (LS) and
  LibriTTS-R (LT-R).}
  \label{tab:rategrid}
\end{table*}

\subsection{Downstream task-transfer panel}
\label{sec:downstream}

The panel contains 11{,}926 clips from LibriSpeech, LibriTTS-R, VCTK, and
EARS, with 23{,}852 speaker-verification trials per condition. dWER is
relative to frozen clean-reference Whisper transcripts; EER uses one pooled
threshold. The two anchors are waveform filters. On this pre-promotion
checkpoint, LILAC is last among the five learned codecs on dWER and
second-worst on EER.

\begin{table}[!htbp]
  \centering
  \footnotesize
  \setlength{\tabcolsep}{5pt}
  \begin{tabular}{lrr}
    \toprule
    Condition & dWER (\%) & Speaker EER (\%) \\
    \midrule
    Anchor, 7.0\,kHz (waveform) & 2.49 & 0.14 \\
    DualCodec @1.23\,kb/s      & 4.78 & 2.11 \\
    Anchor, 3.5\,kHz (waveform) & 4.85 & 3.17 \\
    BigCodec @1.04\,kb/s       & 5.93 & 3.87 \\
    FocalCodec @0.65\,kb/s     & 6.12 & 6.39 \\
    Mimi @$\sim$1.1\,kb/s      & 7.35 & 8.26 \\
    \textbf{LILAC (ours) @0.75\,kb/s} & 9.90 & 7.72 \\
    \bottomrule
  \end{tabular}
  \caption{Single-pass downstream task transfer on the shared four-corpus
  panel.}
  \label{tab:downstream-full}
\end{table}

\paragraph{Speaker EER under repeated re-encoding.}
This panel uses 100 reader-balanced LibriSpeech clips from 40 speakers,
three fixed negative trials per clip, and WavLM-SV x-vectors
(\texttt{microsoft/wavlm-base-plus-sv}; Chen et al., 2022). At the
pre-specified $K{=}10$ comparison, Mimi and FocalCodec cross above LILAC
on both downstream endpoints ($p_{\mathrm{Holm}}{=}0.0008$); EnCodec is
already above LILAC at $K{=}1$, and DAC remains below it at $K{=}10$.
LILAC uses the final 898k flagship and stays at 4.0\% EER at every depth.

\begin{table}[!htbp]
  \centering
  \footnotesize
  \setlength{\tabcolsep}{3pt}
  \begin{tabular}{lrrrrrr}
    \toprule
    & \multicolumn{6}{c}{Pooled speaker-EER (\%) after $K$ self-encodings} \\
    \cmidrule{2-7}
    Codec & $K{=}1$ & 2 & 5 & 10 & 50 & 100 \\
    \midrule
    DAC            & \textbf{0.0} & \textbf{0.0} & \textbf{0.0} & \textbf{3.0} & 30.2 & 46.0 \\
    FocalCodec     & 3.0 & 6.8 & 15.0 & 27.0 & 40.0 & 41.0 \\
    Mimi           & 3.8 & 8.8 & 14.0 & 28.2 & 35.2 & 35.0 \\
    EnCodec        & 10.0 & 12.0 & 12.0 & 12.0 & 12.0 & 12.0 \\
    \textbf{LILAC (ours)} & 4.0 & 4.0 &
    4.0 & 4.0 & \textbf{4.0} & \textbf{4.0} \\
    \bottomrule
  \end{tabular}
  \caption{Speaker-EER under iterated re-encoding; LILAC's row is
  invariant by construction (bit-identical codes at every generation).}
  \label{tab:eer-under-k}
\end{table}

\section{Crowdsourced listening-test details}
\label{sec:supp-mushra}

\begin{table}[!htbp]
  \centering
  \footnotesize
  \begin{tabular}{@{}p{0.23\linewidth}p{0.70\linewidth}@{}}
    \toprule
    Field & Value \\
    \midrule
    Platform & Paid anonymous crowd pool; audio-device flag; \$3.42 per
    session (\$136.62 total); no personally identifying information. \\
    Formal panel & 49 completed training; 40 completed scored sessions;
    all 40 passed screening. \\
    Items & Nine clips, three each from LibriSpeech test-clean,
    LibriTTS-R test, and VCTK validation; 4--10\,s; $-23$\,LUFS; 48\,kHz. \\
    Conditions & Hidden reference; 3.5 and 7\,kHz anchors; LILAC (0.75\,kb/s);
    FocalCodec (0.65); WavTokenizer (0.90); SNAC (0.98); DualCodec (1.23). \\
    Instructions & Play the reference first; score an indistinguishable
    candidate 100 and decrease scores with increasing difference. \\
    Screening & S1: complete $9\times8$ response rectangle. S2: both catch
    slates rank the 3.5\,kHz anchor below hidden and duplicate references.
    S3: nonconstant experimental ratings. \\
    Analysis & Equal-corpus mixed model; listener and clip random intercepts;
    four pre-declared LILAC-minus-peer contrasts; 2{,}000-refit 95\%
    intervals; two-sided Wald tests with Holm correction. \\
    Sensitivity & 10{,}000-draw crossed listener/clip bootstrap; all four
    contrast signs agree. \\
    Protocol status & Screening and analysis were fixed before formal
    collection began; all pre-formal ratings are excluded. Participation
    under the platform's worker agreement; no separate institutional
    ethics review. \\
    \midrule
    \multicolumn{2}{@{}l}{\textbf{Formal-panel diagnostics}} \\
    \multicolumn{2}{@{}c@{}}{%
      \begin{tabular}{@{}lr@{\hspace{3em}}lr@{}}
        Retained sessions & 40/40 &
        Hidden-reference mean & 90.6 \\
        ICC(2,1) & 0.655 &
        Duplicate-reference disagreement, median & 4.5 \\
        Duplicate-reference disagreement, mean & 9.25 &
        Duplicate-reference disagreement, maximum & 50 \\
        Within-listener rating SD, median & 28.6 &
        7\,kHz anchor $>90$ ($>25\%$ of listeners) & 2/9 \\
      \end{tabular}} \\
    \bottomrule
  \end{tabular}
  \caption{Formal crowdsourced listening-test design and panel diagnostics.}
  \label{tab:listening-design}
\end{table}

\section{Displaced Main-Text Tables}
The following tables support prose claims in the main paper: the
stem/augmentation ablation, the speaker-EER/dWER-under-$K$ grid, per-seed
results, the generator loss configuration, and the crowdsourced
MUSHRA-style results.

\begin{table}[!htbp]
\centering
\small
\setlength{\tabcolsep}{4.5pt}
\begin{tabular}{@{}lcccc@{}}
\toprule
 & \multicolumn{2}{c}{LibriSpeech} & \multicolumn{2}{c}{LibriTTS-R} \\
\cmidrule(lr){2-3}\cmidrule(lr){4-5}
Configuration & SCQ$\downarrow$ & dWER$\downarrow$ & SCQ$\downarrow$ & dWER$\downarrow$ \\
\midrule
Neither & 0.531 & 0.124 & 0.572 & 0.121 \\
Stem only & 0.346 & 0.115 & 0.604 & 0.252 \\
Stem + aug. & 0.352 & 0.109 & 0.327 & 0.098 \\
\bottomrule
\end{tabular}
\caption{Anti-imaging stem and target-preserving augmentation ablation
at matched 300k steps.}
\label{stab:ablation-stem-aug}
\end{table}

\begin{table}[!htbp]
\centering
\small
\setlength{\tabcolsep}{3.2pt}
\begin{tabular}{@{}lcccccc@{}}
\toprule
 & \multicolumn{3}{c}{EER (\%)$\downarrow$} & \multicolumn{3}{c}{dWER$\downarrow$} \\
\cmidrule(lr){2-4}\cmidrule(lr){5-7}
Codec & $K{=}1$ & $K{=}10$ & $K{=}100$ & $K{=}1$ & $K{=}10$ & $K{=}100$ \\
\midrule
LILAC & 4.00 & 4.00 & 4.00 & 0.103 & 0.103 & 0.103 \\
FocalCodec & 3.00 & 27.00 & 41.00 & 0.072 & 0.489 & 1.335 \\
Mimi & 3.83 & 28.17 & 35.00 & 0.099 & 0.746 & 1.012 \\
EnCodec & 10.00 & 12.00 & 12.00 & 0.115 & 0.138 & 0.143 \\
DAC & 0.00 & 3.00 & 46.00 & 0.024 & 0.087 & 1.450 \\
\bottomrule
\end{tabular}
\caption{Speaker EER and dWER after $K$ self-encoding cycles
(Figure~4 of the main paper, selected depths). LILAC's values are
identical at every $K$.}
\label{stab:under-k}
\end{table}

\begin{table}[!htbp]
\centering
\small
\setlength{\tabcolsep}{3pt}
\begin{tabular}{@{}lcccccc@{}}
\toprule
 & \multicolumn{3}{c}{LibriSpeech} & \multicolumn{3}{c}{LibriTTS-R} \\
\cmidrule(lr){2-4}\cmidrule(lr){5-7}
Seed & UTMOS$\uparrow$ & dWER$\downarrow$ & SCQ$\downarrow$ & UTMOS$\uparrow$ & dWER$\downarrow$ & SCQ$\downarrow$ \\
\midrule
1 & 3.990 & 0.124 & 0.344 & 4.118 & 0.118 & 0.293 \\
2 & 4.014 & 0.125 & 0.326 & 4.148 & 0.099 & 0.275 \\
3 & 4.009 & 0.122 & 0.335 & 4.164 & 0.092 & 0.283 \\
4 & 4.036 & 0.122 & 0.317 & 4.166 & 0.101 & 0.266 \\
5 & 4.043 & 0.121 & 0.322 & 4.172 & 0.100 & 0.275 \\
\midrule
mean & 4.019 & 0.123 & 0.329 & 4.154 & 0.102 & 0.279 \\
sd & 0.021 & 0.002 & 0.011 & 0.022 & 0.010 & 0.010 \\
\bottomrule
\end{tabular}
\caption{Per-seed results: five seeds of the recipe, each trained to
600k steps and averaged over its final ten checkpoints.}
\label{stab:seed-variance-perseed}
\end{table}

\begin{table}[!htbp]
\centering
\small
\setlength{\tabcolsep}{3.5pt}
\begin{tabular}{@{}llc@{}}
\toprule
Loss term & Configuration & Weight \\
\midrule
Mel reconstruction & FFT $\{512,1024,2048\}$ & 15.0 \\
 & \quad mels $\{64,128,256\}$ & \\
Multi-res.\ STFT & FFT $\{128,\dots,2048\}$, 5 res. & 1.0 \\
Adversarial (hinge) & MPD $\{2,3,5,7,11\}$ + MS-STFT & 1.0 \\
Feature matching & $L_1$, all discriminator features & 2.0 \\
\bottomrule
\end{tabular}
\caption{Generator loss terms. Mel hop lengths are one quarter of each
FFT size; STFT-loss hops are likewise FFT$/4$. Adversarial and
feature-matching terms activate after a 5{,}000-step warm-up ramp.}
\label{stab:loss-config}
\end{table}

\begin{table}[!htbp]
  \centering
  \small
  \setlength{\tabcolsep}{5pt}
  \begin{tabular}{lrrrr}
    \toprule
    System & kb/s & MUSHRA [95\% CI] & $\Delta_{\mathrm{LILAC-peer}}$ [95\% CI] & $p_{\mathrm{Holm}}$ \\
    \midrule
    Hidden reference & --- & 90.6 [85.0, 96.2] & & \\
    DualCodec & 1.23 & 74.8 [69.2, 80.5] & $-23.4$ $[-32.2,\,-15.6]$
    & $<\!10^{-9}$ \\
    7~kHz anchor & --- & 60.0 [54.4, 65.7] & & \\
    SNAC & 0.98 & 59.9 [54.2, 65.5] & $-8.5$ $[-14.3,\,-2.7]$ &
    $7\!\times\!10^{-7}$ \\
    WavTokenizer & 0.90 & 59.6 [54.0, 65.2] & $-8.2$ $[-21.8,\,+5.0]$
    & $1\!\times\!10^{-6}$ \\
    LILAC & 0.75 & 51.4 [45.8, 57.0] & & \\
    FocalCodec & 0.65 & 48.5 [42.9, 54.1] & $+2.9$ $[-3.3,\,+9.0]$ & $0.077$ \\
    3.5~kHz anchor & --- & 29.2 [23.6, 34.8] & & \\
    \bottomrule
  \end{tabular}
  \par\smallskip
  \footnotesize $\Delta$ intervals are individual conservative refit-LMM
  bootstrap 95\% CIs (measured null coverage $0.99$ -- deliberately
  conservative); $p$-values are two-sided LMM Wald tests Holm-adjusted
  across the four pre-declared contrasts. The two procedures differ by
  construction, so a small $p$ can coexist with an interval crossing zero
  (WavTokenizer); both are reported as pre-registered, with no success
  gate.
  \caption{Crowdsourced MUSHRA-style results ($N{=}40$ retained listeners,
    nine items): estimated marginal means with 95\% CIs, and the
    pre-declared LILAC-minus-peer contrasts $\Delta$.}
  \label{stab:mushra}
\end{table}

\fi

\end{document}